\documentclass[aip,reprint]{revtex4-1}

\newcommand{\blot}[1]{}

\newtheorem{theorem}{Theorem}[section]

\newtheorem{proposition}[theorem]{Proposition}

\newcommand{\eqd}{\stackrel{\triangle}{=}}

\usepackage{float}
\usepackage[section]{placeins}
\usepackage{graphicx}% Include figure files
\usepackage{dcolumn}% Align table columns on decimal point
\usepackage{bm}% bold math
\usepackage{ulem,color,mathrsfs,amsmath,amsfonts}

\begin{document}

\preprint{AIP/123-QED}

\title[Distributional fixed point equations for island nucleation in one dimension]{Distributional fixed point equations for island nucleation in one dimension: a retrospective approach for capture zone scaling}
%\thanks{Footnote to title of article.}

\author{P. A. Mulheran}%
 \email{paul.mulheran@strath.ac.uk}
 \affiliation{Department of Chemical and Process Engineering, University of Strathclyde, Glasgow}

\author{K. P. O'Neill}
 \email{kenneth.o-neill@strath.ac.uk}
 \affiliation{Department of Mathematics and Statistics, University of Strathclyde, Glasgow}
 
\author{M. Grinfeld}
 \email{m.grinfeld@strath.ac.uk}
 \affiliation{Department of Mathematics and Statistics, University of Strathclyde, Glasgow}
 
 \author{W. Lamb}
  \email{w.lamb@strath.ac.uk}
  \affiliation{Department of Mathematics and Statistics, University of Strathclyde, Glasgow}
 

%\affiliation{ 
%Authors' institution and/or address%\\This line break forced with \textbackslash\textbackslash
%}

% \homepage{http://www.Second.institution.edu/~Charlie.Author.}
%\affiliation{%
%Second institution and/or address%\\This line break forced% with \\
%}%

\date{\today}% It is always \today, today,

%Use showkeys class option if keyword

\begin{abstract}

The distributions of inter-island gaps and captures zones for islands nucleated on a one-dimensional substrate during submonolayer
deposition are considered using a novel retrospective view. This provides an alternative perspective on why scaling occurs in this 
continuously evolving system. Distributional fixed point equations for the gaps are derived both with and without a mean field 
approximation for nearest neighbour gap size correlation. Solutions to the equations show that correct consideration of 
fragmentation bias justifies the mean field approach which can be extended to provide closed-from equations for the capture zones. 
Our results compare favourably to Monte Carlo data for both point and extended islands using a range of critical island size 
$i=0,1,2,3$. We also find satisfactory agreement with theoretical models based on more traditional fragmentation theory approaches.

\end{abstract}			      
			      
\maketitle

%Section 1 - Introduction

\section{INTRODUCTION\label{Sec1_Intro}}

Scale invariance during the nucleation and growth of islands driven by
monomer deposition is an intriguing phenomenon \cite{AF95}. Island
size distributions, and the distribution of capture zones which
underlie the island growth rates, evolve towards scaling forms despite
on-going nucleation of new islands with the concomitant disruption to
the existing capture zones \cite{PAM04}. The form of the scaling
functions depends on the critical island size $i$, where $i+1$ is the
smallest stable island size. A number of theoretical approaches have
been used to model this behaviour, ranging from mean field models
which neglect the variation in capture zone sizes \cite{Venables73,
  BE92, Amar94, BC94, BW91, Ratsch94} due to spatial arrangements of
the islands, to those which attempt to include this information
explicitly \cite{MB95, BE96, MR00, Amar01, EB02}. All these approaches
can be characterised as forward-looking in the sense that they are
based on predicting how size distributions evolve as new islands
nucleate. 

Recently, for island nucleation and growth during submonolayer deposition, 
Pimpinelli and Einstein introduced a new theory for the capture zone distribution (CZD), 
employing the Generalised Wigner Surmise (GWS) \cite{PE07},

\begin{equation}
 \label{GWS}
 P(s) = a_\beta s^\beta \exp(-b_\beta s^2),
\end{equation}
where $a_{\beta}$ and $b_{\beta}$ are normalising constants, and

\begin{equation}
 \label{eqGWS_beta}
 \beta \;
  =
  \left\{
  \begin{array}{cc}
  \frac{2}{d}(i+1) \ \mbox{ if } d = 1,2 \\
  i+1 \ \ \ \mbox{ if } d=3.
  \end{array}
  \right.
\end{equation}
Based on excellent visual comparisons between the GWS and Monte Carlo (MC) simulation data taken from the 
literature \cite{PE07}, the GWS has already been explored further \cite{OR11} and its functional form questioned   \cite{Li10}.
 For example, Shi {\it et. al.} \cite{Shi09} studied $i=1$ point-island models in dimensions $d=1,2,3,4$. By investigating the peak of the simulated CZD, 
Shi {\it et. al.} find that the CZD is more sharply peaked and narrower than the GWS suggests, and a better choice of $\beta$ is $3$ rather than $\beta=2$ 
for $d=2,3$. Moreover, for $d=1$, it is notable that the peak height analysed by Shi {\it et. al.} suggests that the predicted value of $\beta=4$ is not correct.
 
In [\onlinecite{PE07}], the island nucleation rate 
is discussed in terms of the monomer density $n$, and the probability of $(i+1)$ monomers coinciding is used to give the nucleation rate as 
$n^{i+1}$ . This is the same physical basis Blackman and Mulheran have used for their fragmentation theory in the $i=1$ case to investigate 
the gap size distribution (GSD) and, subsequently, the CZD \cite{BM96}. This motivated our recent works, which we discuss next.

In [\onlinecite{GLOM11}], we have extended the analysis of the original fragmentation equations \cite{BM96} to the case of general $i\ge 0$. 
We have been able to derive the small- and large-size asymptotics of the GSD, and by assuming random mixing of the gaps caused by the nucleation
 process, we have also derived the small-size asymptotics for the CZD for general $i$ and the large-size behaviour for $i=0$. One key feature to 
emerge from the fragmentation equations is that the asymptotic behaviour of the CZD is different to that of the GWS \cite{PE07}. In addition to this, 
recent work by Gonz\'{a}lez {\it et. al.} \cite{GPE11} has revisited the $i=1$ case, developing the original fragmentation equation \cite{BM96} and GWS arguments 
in response to deviations between prediction and simulation. In our recent work \cite{OGLM12} we explored simulation results for the one-dimensional 
(1-D) model with $i=0,1,2,3$, and considered the relative merits of the GWS \cite{PE07} and the fragmentation theory \cite{GLOM11} approaches. 
The paper \cite{OGLM12} concludes that the GWS predictions for the small-size CZD scaling work well since they bisect
 the exponents from the alternative nucleation mechanisms. As discussed elsewhere \cite{GLOM11}, the predicted formula for the parameter 
$\beta$ of the GWS can be brought into line with either nucleation mechanism following the arguments of Pimpinelli and Einstein \cite{PE-reply}. 
Nevertheless, the original prediction of these authors, Eqn.~(\ref{eqGWS_beta}), does provide a convenient point of comparison for our own work in this paper, 
notwithstanding the aforementioned debate over its precise functional form.

The conceptual basis of these and similar works that employ fragmentation theory is one of forward propagation in time of the GSD and CZD. In this paper we shall present an alternative,
retrospective, perspective where we ask how the capture zones present
in the system came to be created. This approach was inspired by Seba \cite{Seba07} who 
investigated a 1-D model aimed at describing the spacing distribution 
between cars parked in an infinitely long street. Seba derived the distributional fixed point equation (DFPE) 

\begin{displaymath}
  X_d \eqd a(1+X_d).
\end{displaymath}
where $X_d$ is the distance between two parked cars, $a$ is an independent random variable with 
a probability density $f(a)$, and the symbol $\eqd$ means that the left- and right-sides of the 
above DFPE have the same distribution. In this paper, we will apply a similar approach to the nucleation of point islands in a 1-D
system. This model allows for a more complete analysis than one with more realistic extended islands, but as we shall show below, 
there is good simulation evidence to suggest that the analysis can equally apply to the more realistic system and is not limited to
our point-island model. We will also compare our results with those from a more
traditional fragmentation theory approach \cite{BM96, GLOM11} as well as the GWS. Our
new perspective provides interesting insight into why scaling occurs
and compares well with simulation data.

% Section -- MC Simulation

\section{MONTE CARLO SIMULATIONS\label{Sec2_MC}}

Island nucleation and growth is widely studied using Monte Carlo (MC)
simulation. A point island approximation is often used both for
clarity and because it approximates the growth of small,
well-separated islands \cite{BE92}, and 1-D systems occur
experimentally during island growth at substrate steps. Here we employ
a 1-D model \cite{BM96} where monomers are deposited at random onto an
initially empty lattice at a deposition rate of $F$ monolayers per
unit time. The monomers diffuse at rate $D$ on the lattice, nucleating
immobile point islands when $i+1$ monomers coincide at a lattice
site. Once nucleated, the islands grow by absorbing any monomers that
hit them; point islands only occupy one lattice site. 
Alternatively, extended island are allowed; such islands grow by capturing monomers 
that diffuse to their edges. Here, extended islands are 1-D structures, so that an
extended island of size $j$ occupies $j$ sites on the lattice. 
When sufficient islands have been nucleated, the 
most likely fate of a deposited monomer is to become absorbed by an existing island 
rather than being incorporated into a new island. It is in this aggregation regime 
of growth where scale invariance is found; note however that island nucleation 
continues still, albeit at a slow rate compared to monomer adsorption.

Since we assume that any monomer cannot evaporate from the substrate, the 
deposition process can be measured by the nominal substrate coverage,
 $\theta=Ft$; in other words $\theta$ is deposition rate times elapsed time. 
For the extended-island model, $\theta$ is a natural measure of substrate coverage,
whereas for point islands it is a convenient measure of time.
The value of $\theta$ for which the aggregation regime (where scale-invariance is found) 
starts is dependent on $i$ and the ratio $R=D/F$; we check that the values for
 $\theta$ are sufficiently high to ensure that we are in the aggregation regime.

Our simulations \cite{OGLM12} were performed on lattices with $10^6$ sites, with 
$R=8\times 10^6$ up to coverage $\theta=100$\%, averaging results over 100 runs. 
For $i=0$ we set the spontaneous nucleation probability to $p_n=10^{-7}$. We use
this data below to validate our theory development.

\begin{figure}[!h]
 \begin{center}
  \mbox{
        \begin{minipage}{1.3in}
        \scalebox{0.20}{\includegraphics{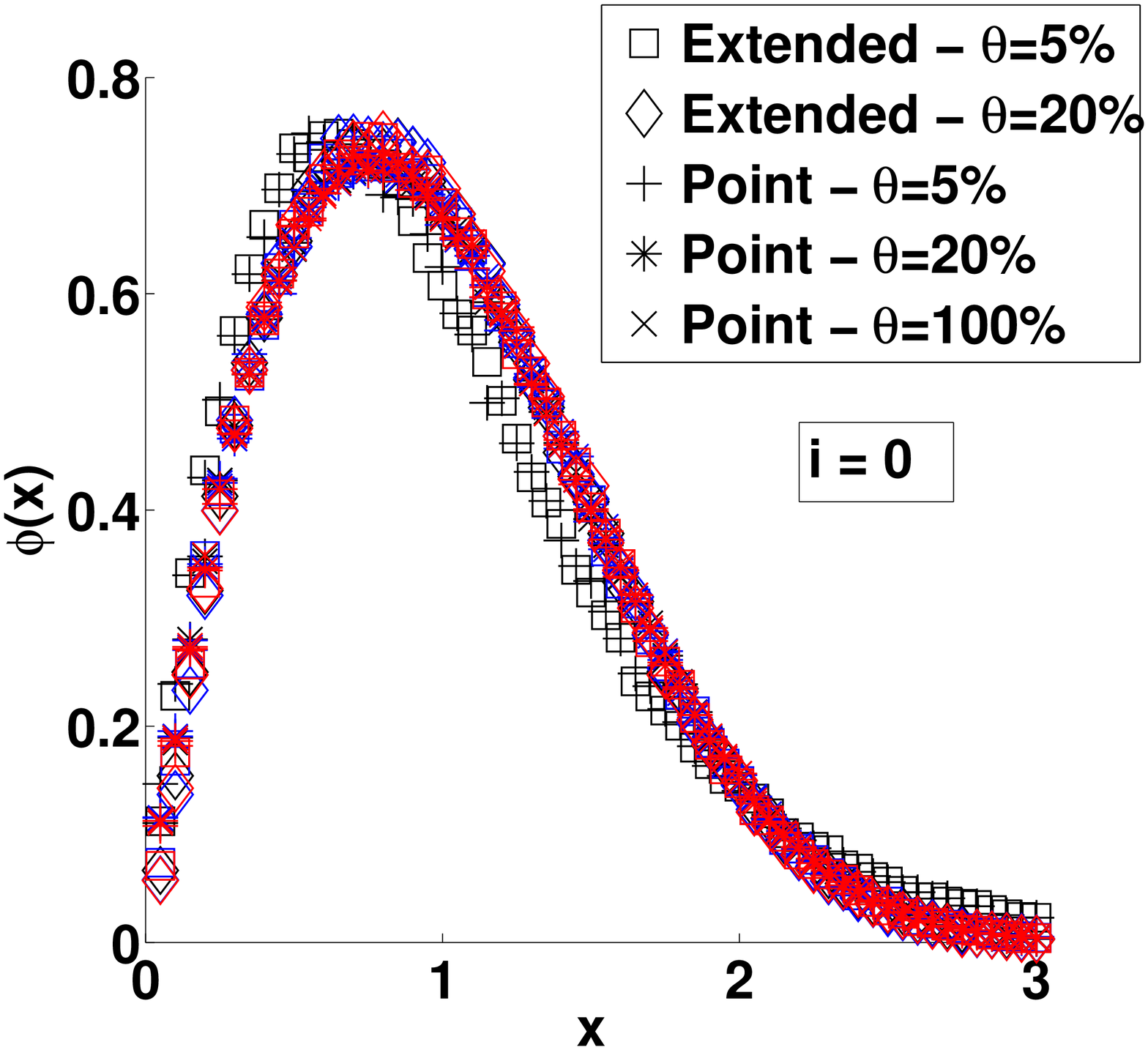}}\\
	    \end{minipage}
	    \qquad \qquad   %Note here, double qquad commands
	    \begin{minipage}{1.3in}
        \scalebox{0.20}{\includegraphics{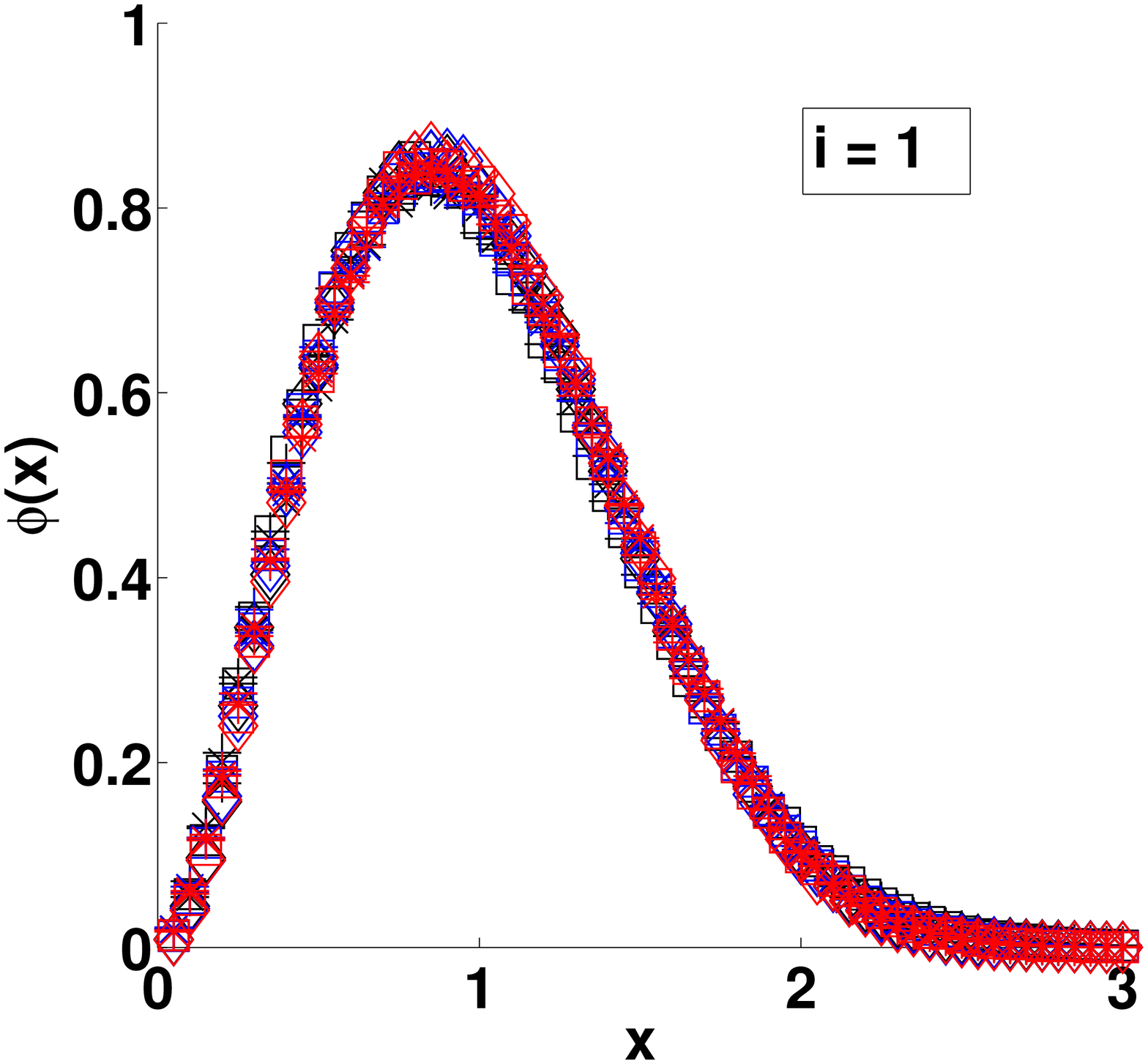}}\\
	    \end{minipage}
        }\\
  \end{center}
  \begin{center}
   \mbox{
        \begin{minipage}{1.3in}
        \scalebox{0.20}{\includegraphics{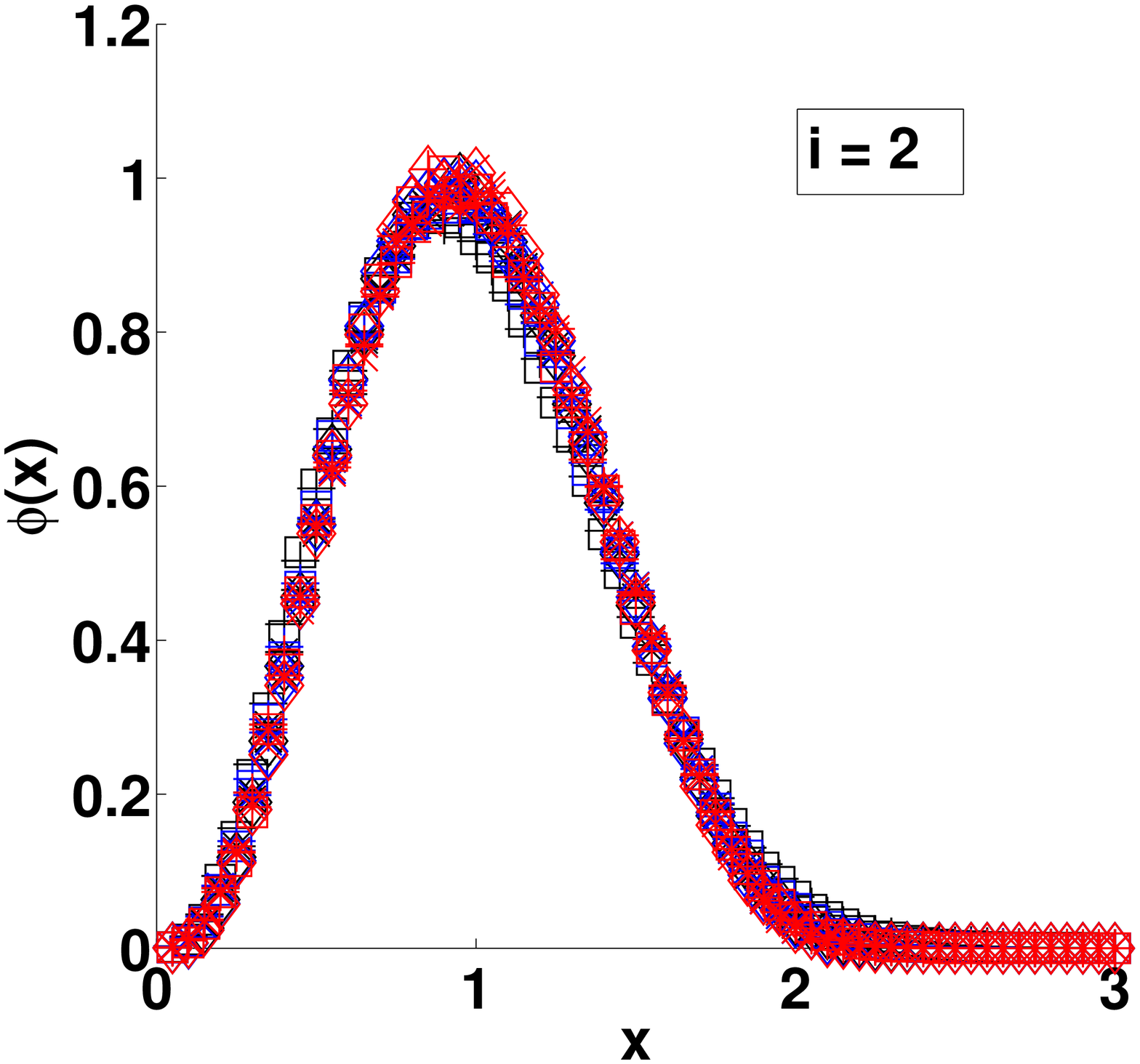}}\\
	    \end{minipage}
	    \qquad \qquad  
	    \begin{minipage}{1.3in}
        \scalebox{0.20}{\includegraphics{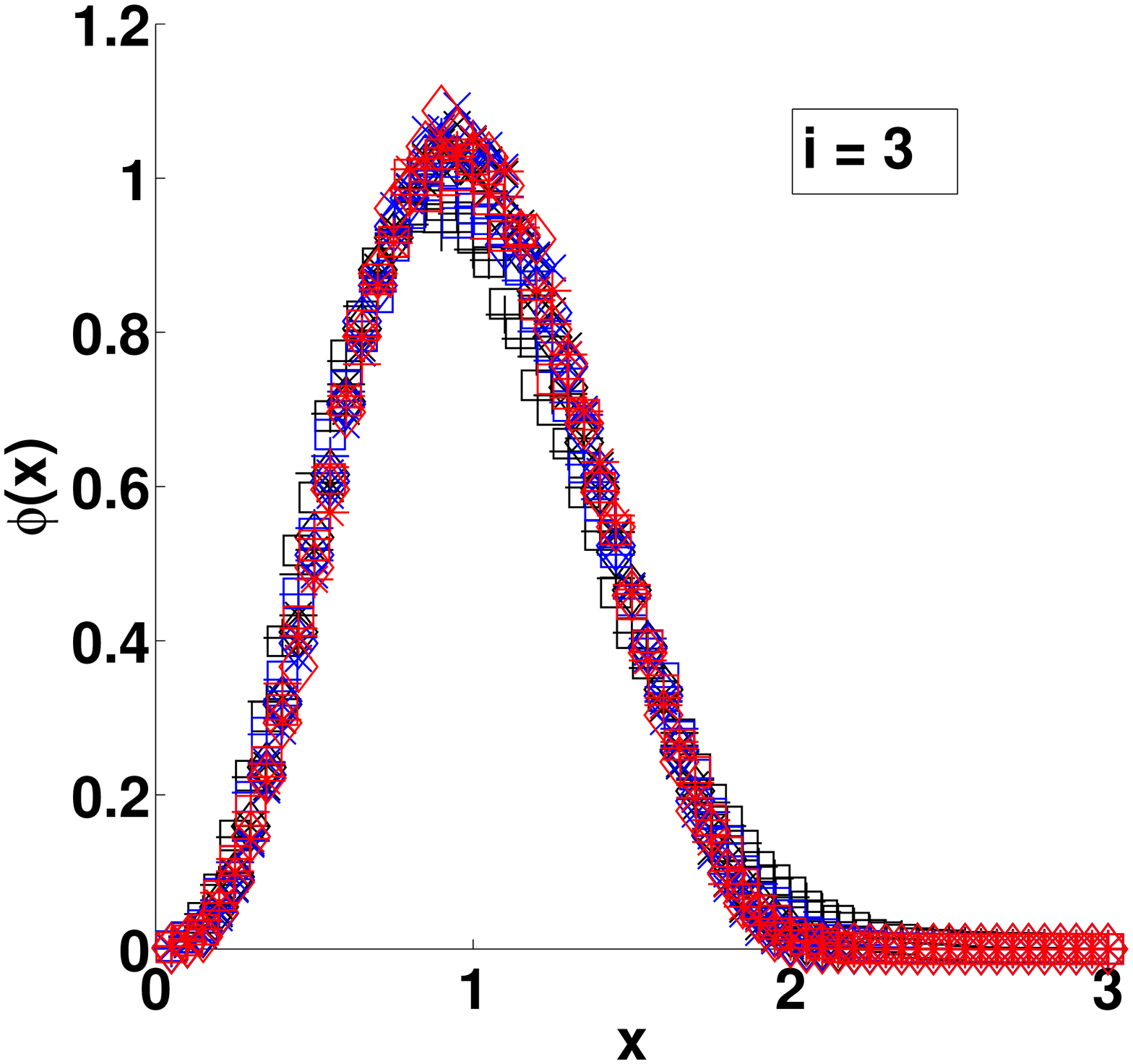}}\\
	    \end{minipage}
       }\\
 \end{center}
\caption{\label{Different_R_cov_GSD} The scaled GSD $\phi$ for extended and point islands with 
$i=0$, $1$, $2$, $3$ and $\theta=5$\%, $20$\% and, in the point-island case only, $100$\% obtained from 
MC simulations [$R=10^7$ (black), $R=10^8$ (blue) and $R=10^9$ (red)]. Note that the data for $R=10^9$
with $i=3$ and $\theta=100$\% are not included.}
\end{figure}

\begin{figure}[!h]
 \begin{center}
  \mbox{
        \begin{minipage}{1.3in} 
        \scalebox{0.20}{\includegraphics{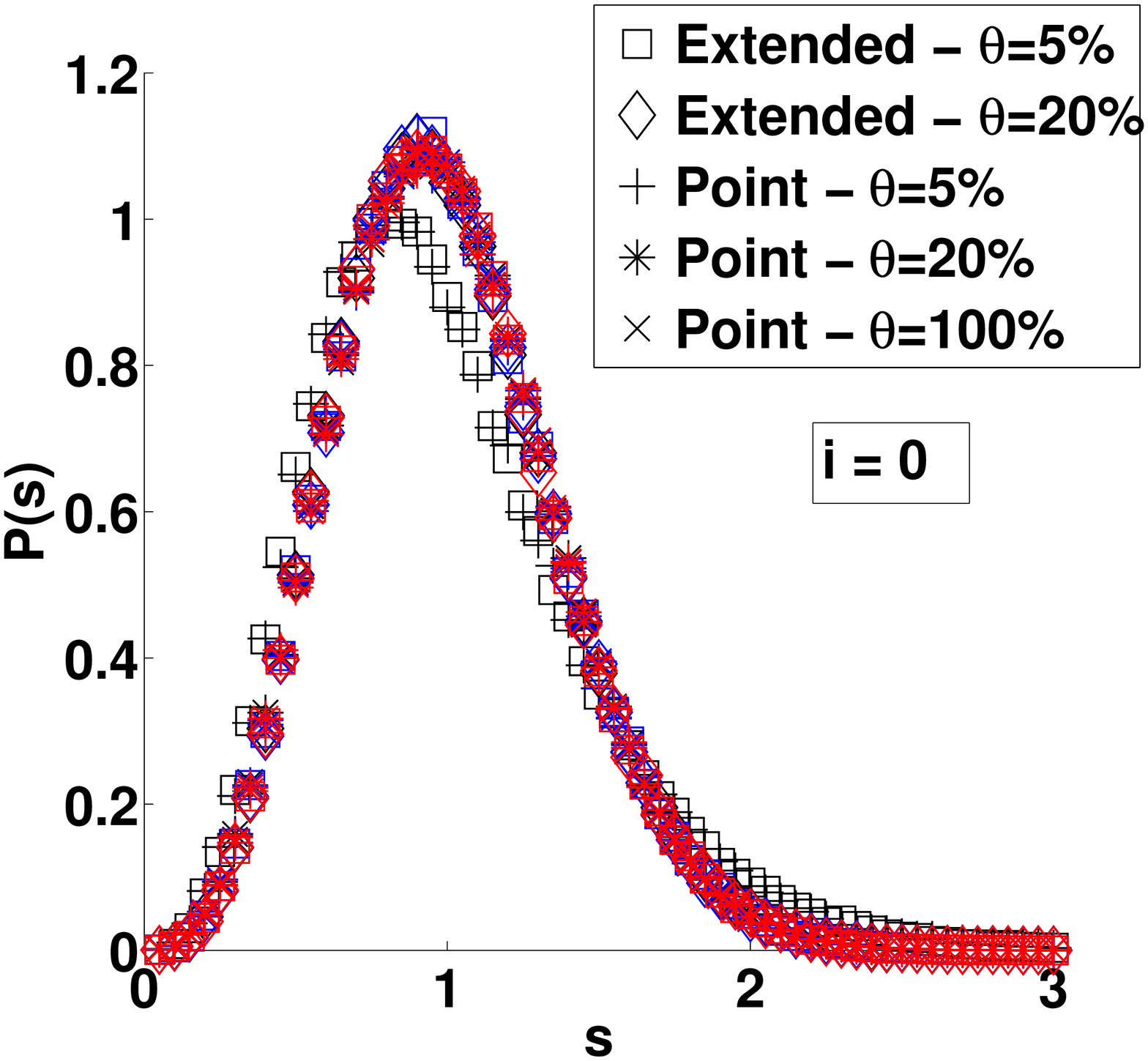}}\\
	    \end{minipage}
	    \qquad \qquad   %Note here, double qquad commands
	    \begin{minipage}{1.3in}
        \scalebox{0.20}{\includegraphics{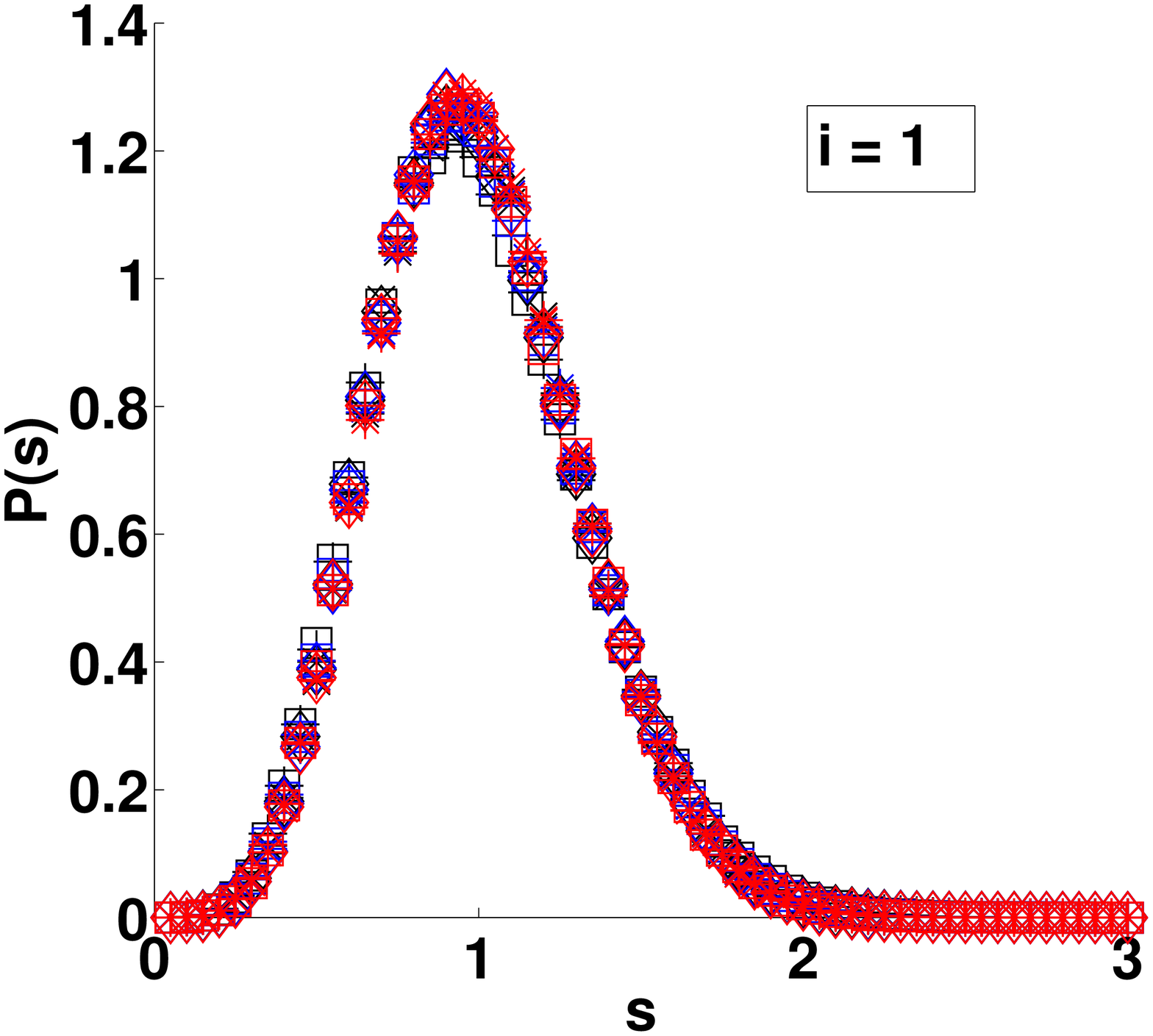}}\\
	    \end{minipage}
        }\\
  \end{center}
  \begin{center}
   \mbox{
        \begin{minipage}{1.3in}
        \scalebox{0.20}{\includegraphics{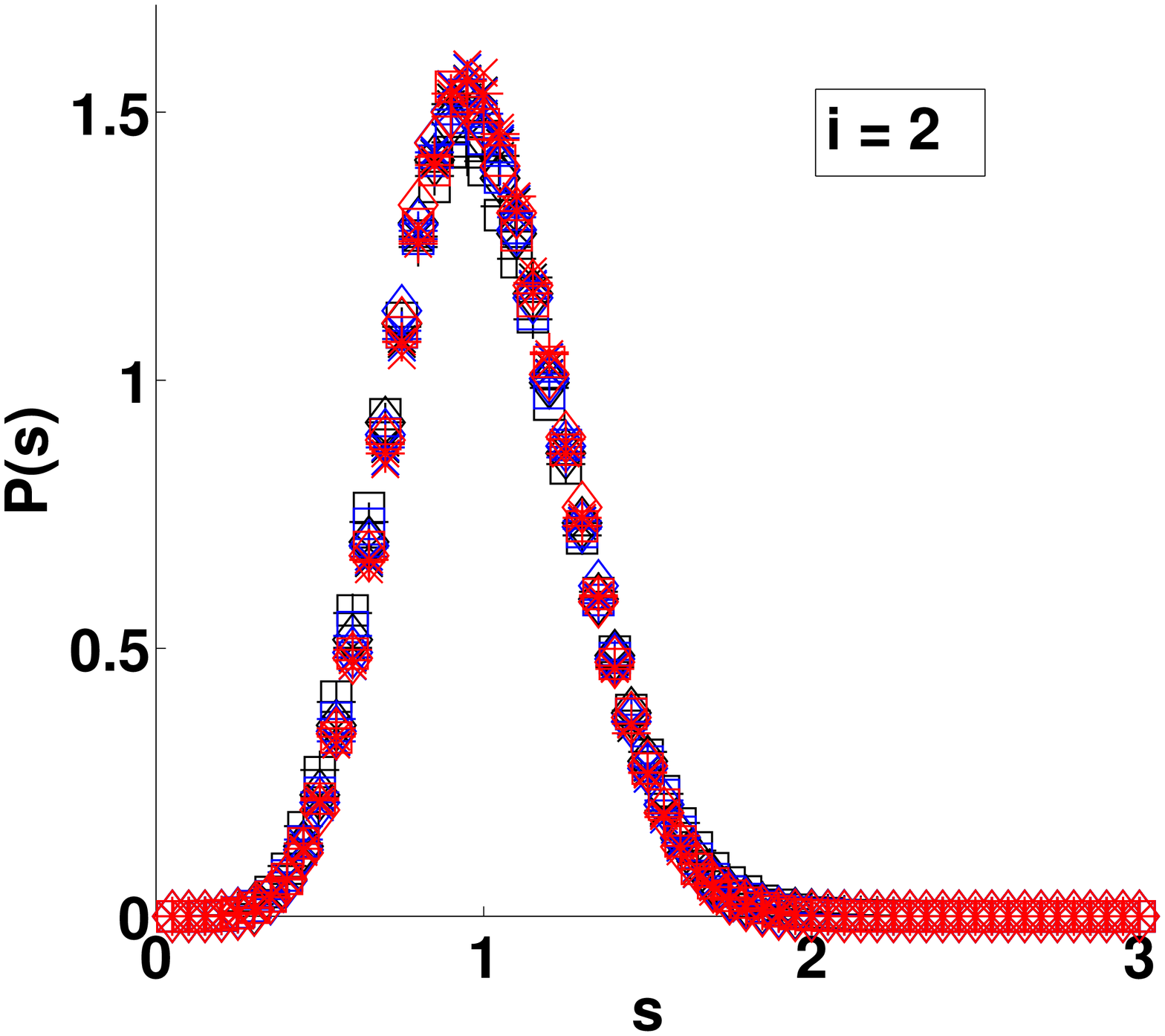}}\\
	    \end{minipage}
	    \qquad \qquad   %Note here, double qquad commands
	    \begin{minipage}{1.3in}
        \scalebox{0.20}{\includegraphics{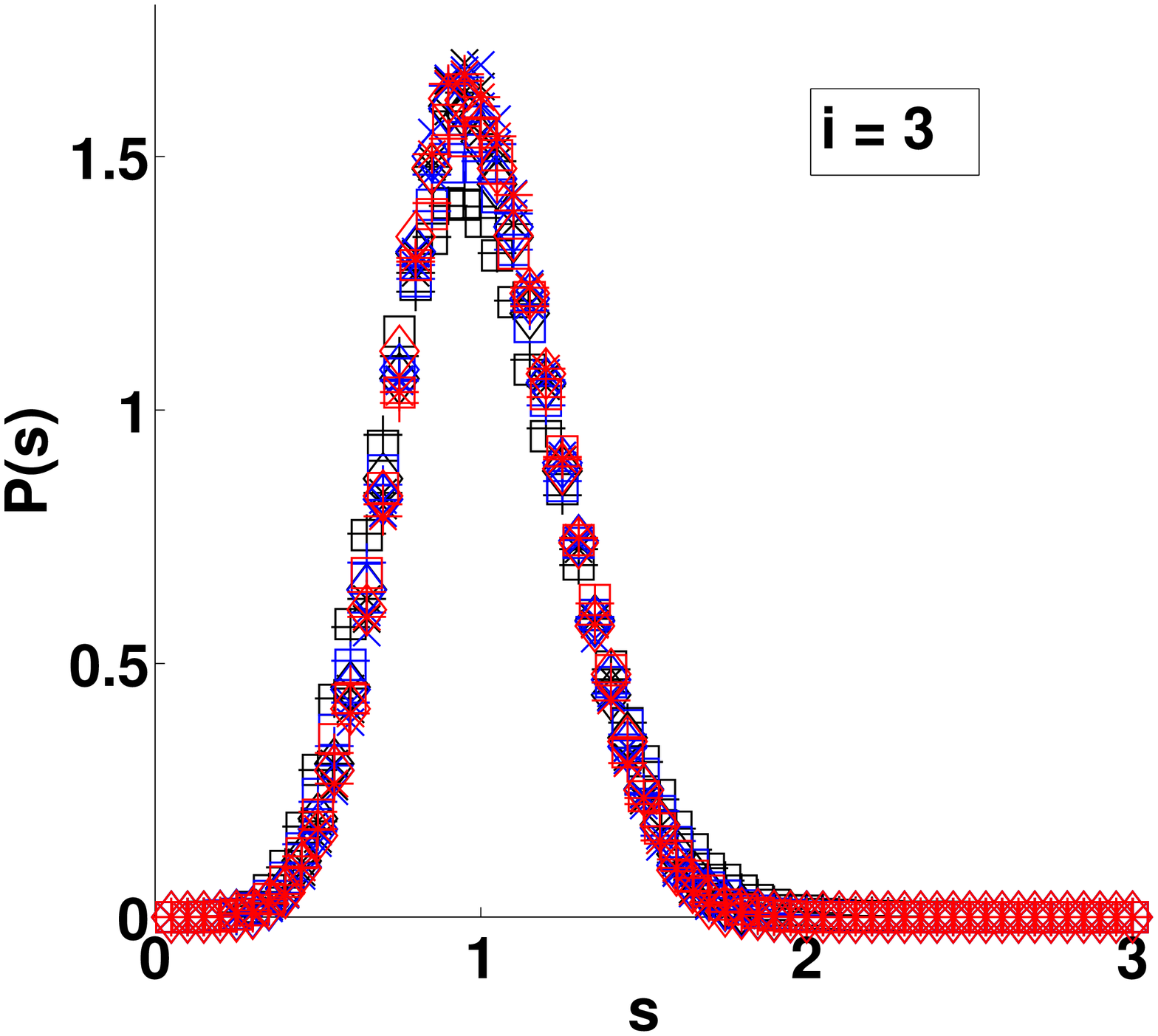}}\\
	    \end{minipage}
        }\\
 \end{center}
\caption{\label{Different_R_cov_CZD} The scaled CZD $P$ for extended and point islands with 
$i=0$, $1$, $2$, $3$ and $\theta=5$\%, $20$\% and, in the point-island case only, $100$\% obtained from 
MC simulations [$R=10^7$ (black), $R=10^8$ (blue) and $R=10^9$ (red)].}
\end{figure}

In addition to this, though it is repeatedly reported in the literature \cite{BE92,BM96,Shi09,OR11} 
that scale-invariance in the island size distribution (ISD), GSD and CZD is observed for large enough $R$,
 it is useful to first consider the dependence of the GSD and CZD on $i$, $R$ and $\theta$ as shown in 
Figures~\ref{Different_R_cov_GSD} and \ref{Different_R_cov_CZD}.
For extended and point islands, we confirm excellent scale-variance for $R>=10^7$ with
various values of $i$. Note that the data for $R=10^7$ at $\theta=5$\% is slightly different from the rest, 
since the aggregation regime occurs at higher coverage for this value of R.
More importantly, we also confirm that the scaled GSD and CZD for the point-island model is
similar to those for the extended islands. Therefore, the point-island model is a very good approximation of the
extended islands at low coverages, i.e. $\theta \le 20$\%.

% Different R and coverages for the GSD

% Different R and coverages for the CZD

As is apparent from these results, we are a long way short of the limit where the 1-D substrate becomes saturated
with islands. This limit is particularly problematic for point island models, since scaling breaks down as 
$\theta \rightarrow \infty$  and the CZD becomes singular \cite{Ratsch05}. Note that for point islands $\theta$ can 
be greater than $100$\% whilst most of the substrate remains free of point islands, since they occupy a single site 
regardless of size. However we have been careful to ensure that we are far from this limit when we use simulation 
data to assess theoretical results below.

% Section -- The MF DFPE for the GSD

\section{THE MEAN-FIELD DFPE APPROACH FOR THE GAP SIZE DISTRIBUTION\label{Sec3_MF_GSD}}

\begin{figure}[!h]
 \begin{center}
  \mbox{
        \begin{minipage}{2.5in}
        \scalebox{0.23}{\includegraphics{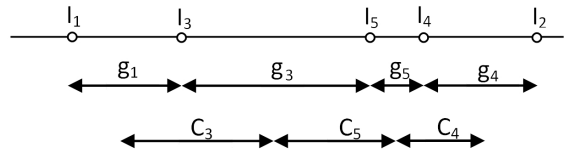}}\\%
	\caption{\label{1D_model} The islands numbered $I_1-I_5$ on
          the one dimensional substrate. The gaps between the islands
          are labelled $g_1$, $g_3$, $g_5$ and $g_4$, and the capture
          zones of islands $I_3$, $I_5$, $I_4$ are labelled $C_3$,
          $C_5$ and $C_4$ respectively.}
	\end{minipage}
        }\\
 \end{center}
\end{figure}

Figure~\ref{1D_model} shows some islands on the lattice, numbered
according to their chronological age, along with their capture zones
$C_3$, $C_4$ and $C_5$. Island $I_3$ has the capture zone of size
$C_3=(g_1+g_3)/2$, where $g_1$ and $g_3$ are the inter-island gaps to
the left and right of $I_3$ respectively. $C_3$ represents the average
growth rate of $I_3$, since any monomers deposited into $C_3$ are more
likely to diffuse to $I_3$ than its neighbours $I_1$ and $I_5$.

Referring to Figure~\ref{1D_model}, let us ask how the inter-island
gap $g_3$ was created. It was formed by the nucleation of the youngest
island in the picture, $I_5$, which occurred in the gap of size
$(g_3+g_5)$ between islands $I_3$ and $I_4$. Generalising, we will
suppose that any randomly chosen gap with size $x$ (scaled to the
average) in the system will have arisen by the fragmentation of a
larger gap formed by combining the gap size of $x$ with a neighbouring gap of size
$y$. In general we do not have the benefit of the chronological ages
to guide us, so we make a mean field (MF) approximation for the size of the
neighbouring gap, namely $y = 1$. Denoting the probability of
fragmenting a gap into proportions $a$ and $(1-a)$ by $f(a)$, we find
the following distributional fixed point equation (DFPE) for the
probability distribution function $\phi(x)$ of gaps $x \in [ 0, \infty)$:

\begin{equation}
 \label{eqDFPE_GSD}
 x \eqd a(1+x).
\end{equation}

This convenient notation (exploited below) states that the
distribution of the variates on the left is equal to that on the right
\cite{Seba07}. Note that we have arrived at the same DFPE that Seba employed 
in his car-parking problem \cite{Seba07}, as discussed in Section~\ref{Sec1_Intro} above. 
As in [\onlinecite{PW}], the DFPE leads to the Integral Equation (IE) for $\phi(x)$,

\begin{equation}
 \label{eqIE_GSD}
 \phi(x) = \int_0^{\min (x,1)} \phi \left( \frac{x}{a}-1 \right)
 \frac{f(a)}{a} da,
\end{equation}
where the derivation of (\ref{eqIE_GSD}) can be found in Appendix~\ref{App.1}. Equation~(\ref{eqDFPE_GSD}) states that
the statistical distribution of gaps is unchanged by the fragmentation of all the gaps incremented 
in scaled size by one. Note that we neglect long-range chronological effects
here, of the type apparent in Figure~\ref{1D_model} for the creation
of gap $g_4$ which arose from the nucleation of island $I_4$ and the
fragmentation of gap $(g_3+g_5+g_4)$. We will return to this point
below.

In the aggregation regime, the probability $f(a)$ of fragmenting a gap
into proportions $a$ and $(1-a)$ is found from the steady-state
monomer density profile \cite{BM96}:

\begin{equation}
 \label{eqProb_GSD}
 f(a) = \frac{a^{\alpha}(1-a)^{\alpha}}{B(\alpha+1,\alpha+1)} = \frac{(2\alpha+1)!}{(\alpha!)^2}a^{\alpha}(1-a)^{\alpha}.
\end{equation}
Here $B(m,n)=\Gamma(m)\Gamma(n)/\Gamma(m+n)$ is the Beta function and $\alpha \in \mathbb{N}$
reflects the dominant nucleation mechanism. For nucleation triggered
by deposition of monomers, $\alpha=i$ for $i=1,2,3\ldots$, whereas for
nucleation resulting from the diffusion of mature monomers
$\alpha=i+1$, $i=0,1,2,\ldots$ \cite{OGLM12}. At the asymptotic limit of 
large $R=D/F$ where $\theta=Ft$, the diffusion mechanism will dominate 
that of deposition. However, in practice, one cannot get to this limit 
in simulations or experiments (nor can we get to $t \rightarrow \infty$). 
Therefore, it is still valid to consider the behaviour whether one mechanism
 or the other dominates because this provides a good bracket to understand our MC data.

\begin{figure}[!h]
 \begin{center}
  \mbox{
        \begin{minipage}{2.5in}
        \scalebox{0.3}{\includegraphics{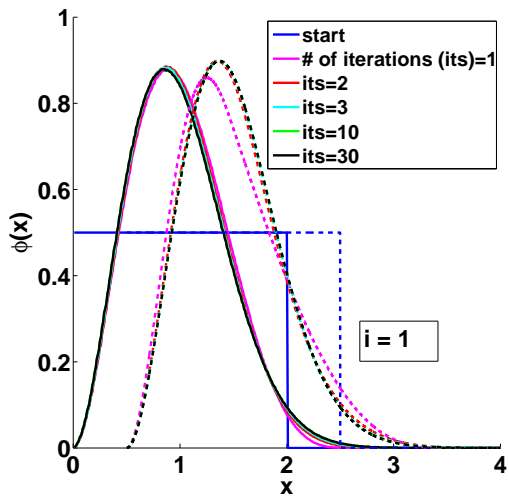}}\\%
	\caption{\label{i1_iterations} The evolution of gap size
          distribution under iteration of equation~(\ref{eqIE_GSD})
          with $i=1$. The solid lines are for $\alpha=i+1$ in
          equation~(\ref{eqProb_GSD}), and the broken lines for
          $\alpha=i$, where the broken lines are shifted along the
          abscissa for clarity.}
	\end{minipage}
        }\\
 \end{center}
\end{figure}

In Figure~\ref{i1_iterations}, by using an iteration scheme of the form

\begin{displaymath}
 \phi_{n+1}(x)=F(\phi_n(x)),
\end{displaymath}
where

\begin{displaymath}
 F(\phi) = \int_0^{\min(x,1)} \phi \left( \frac{x}{a}-1 \right) \frac{f(a)}{a} \ da,
\end{displaymath}
we show the convergence of iterates of equation~(\ref{eqIE_GSD}) 
starting from a rectangular distribution, with $f(a)$ given by 
equation~(\ref{eqProb_GSD}). The limit satisfies the 
DFPE~(\ref{eqDFPE_GSD}), and so is the form that we wish to 
compare to the scale-invariant GSD found in the MC simulations.

\begin{figure}[!h]
 \begin{center}
  \mbox{
        \begin{minipage}{1.3in}
        \scalebox{0.20}{\includegraphics{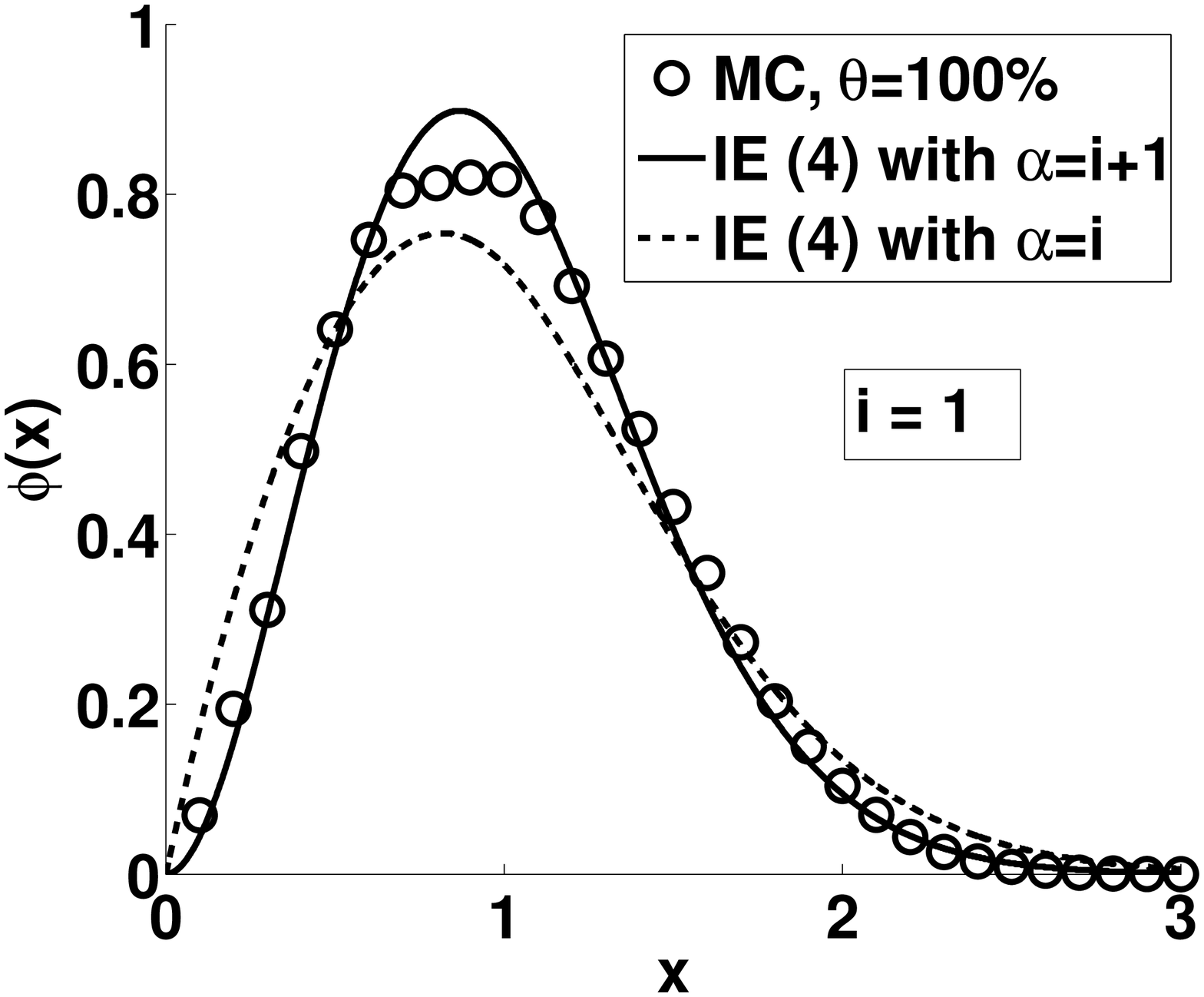}}\\
	    \end{minipage}
	    \qquad \qquad   %Note here, double qquad commands
	    \begin{minipage}{1.3in}
        \scalebox{0.20}{\includegraphics{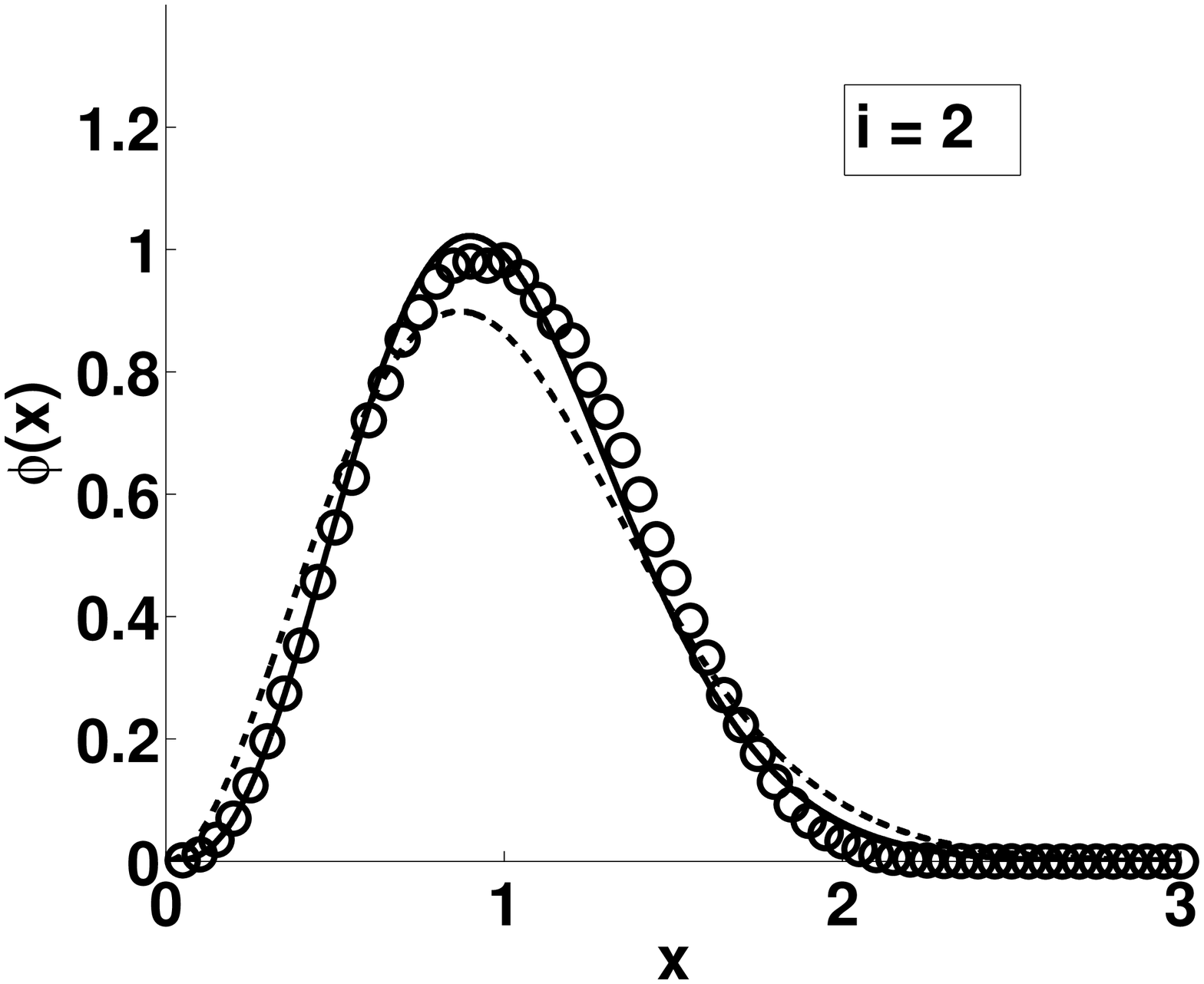}}\\
	    \end{minipage}
        }\\
  \end{center}
  \begin{center}
   \mbox{
        \begin{minipage}{1.3in}
        \scalebox{0.20}{\includegraphics{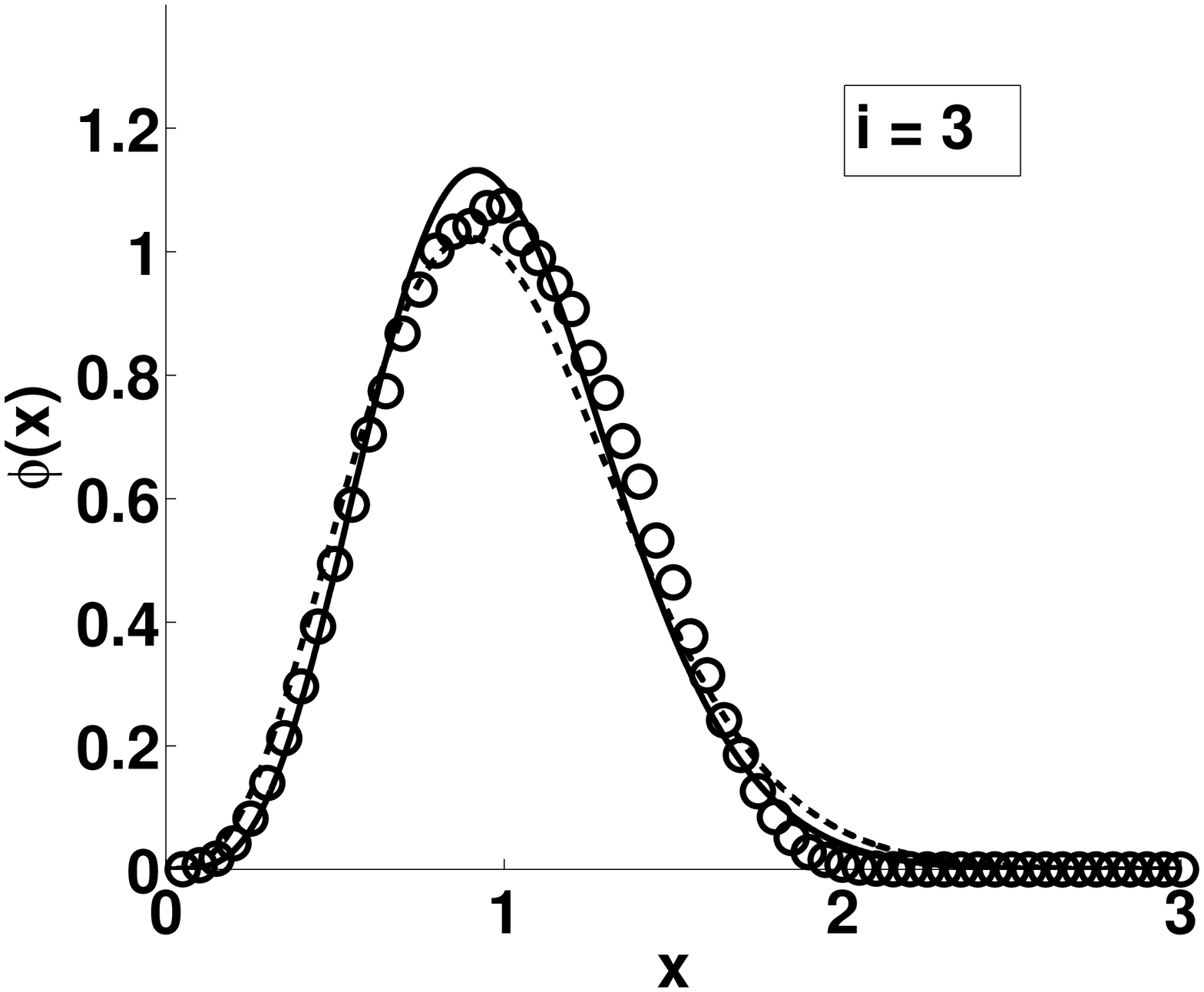}}\\
	    \end{minipage}
	    \qquad \qquad   %Note here, double qquad commands
	    \begin{minipage}{1.3in}
        %\scalebox{0.22}{\includegraphics{i0_IE_GSD_cov100_dfpepaper.eps}}\\
        \scalebox{0.20}{\includegraphics{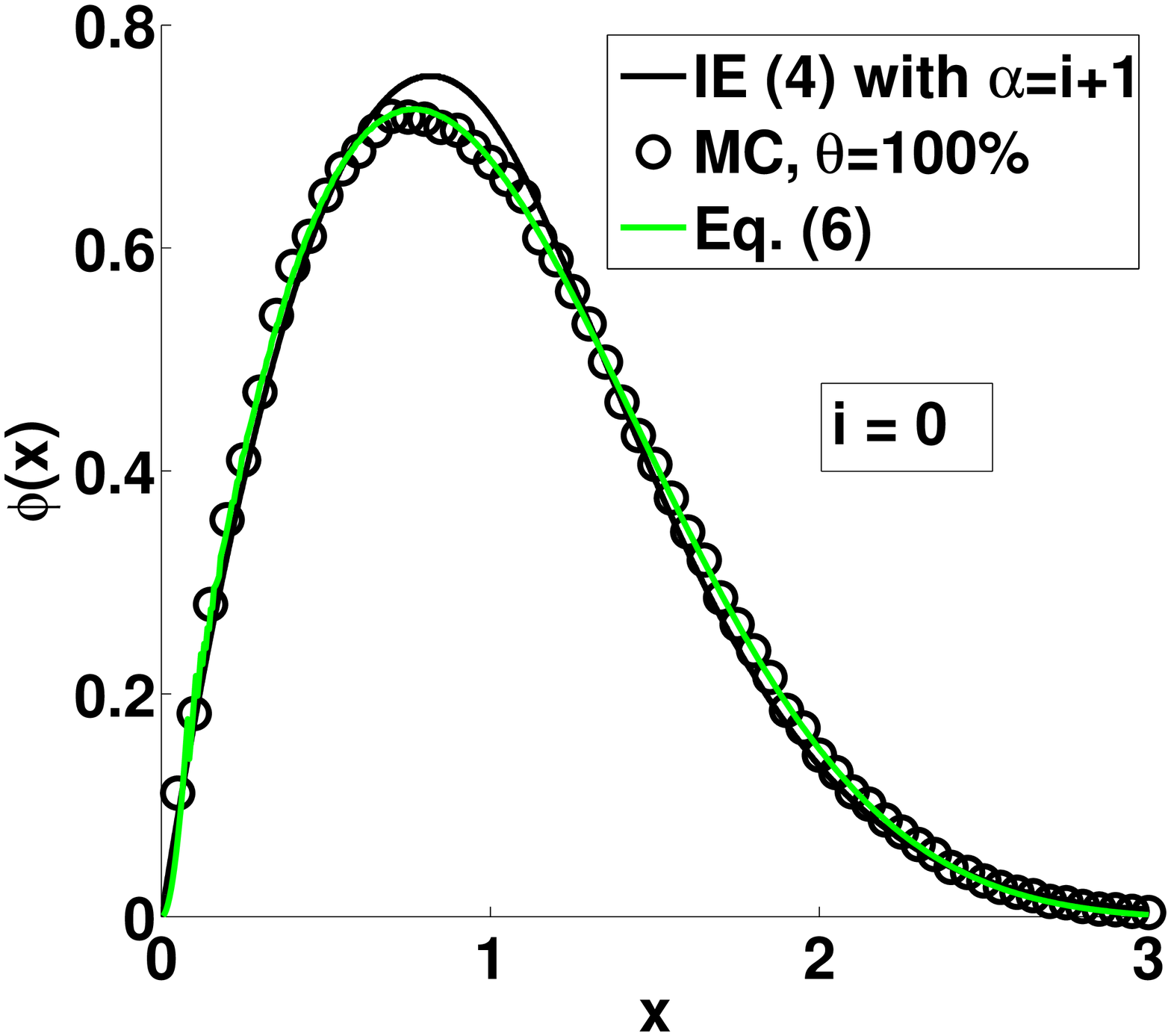}}\\
	    \end{minipage}
        }\\
 \end{center}
\caption{\label{i_0_1_2_3_IE_GSD} The GSDs compared to histograms of
  MC data \cite{OGLM12}, i.e. $R=8\times 10^6$, for various critical island size $i$, taken at
  nominal coverage $\theta=100$\% . The solid curves are the converged solutions to
  equation~(\ref{eqIE_GSD}) with $\alpha=i+1$ with $i=0,1,2,3$, and
  the broken lines are for $\alpha=i$ with $i=1,2,3$.}
\end{figure}

In Figure~\ref{i_0_1_2_3_IE_GSD} we compare the numerically computed
fixed points of equation~(\ref{eqIE_GSD}), which we denote by
$\phi_{\alpha}(x)$, with the GSDs of our MC simulations \cite{OGLM12}
for various critical island size $i$. The comparison is rather
good. For $i=1,2,3$, we see that the observed GSD lies between that of
the $\alpha=i$ and $\alpha=i+1$ distributional fixed point
solutions. This can be expected since we have found elsewhere that
island nucleation is driven by both deposition events and purely
diffusional fluctuations in monomer density \cite{OGLM12}. For
spontaneous nucleation where $i=0$, only the $\alpha=i+1=1$ model is
physically reasonable, since there is no possibility of a monomer
depositing close to a pre-existing critical island of size $i$ in this
case.

It is interesting to ask how the solutions to the DFPE
equation~(\ref{eqIE_GSD}) compare to those of the forward-propagated
fragmentation theory equations, for which the asymptotic behaviours
are known \cite{BM96, GLOM11}. Equation~(\ref{eqIE_GSD}) can be
rewritten as

\begin{displaymath}
\phi(x)= \int_{\max (0,x-1)}^\infty \phi(s)  f\left(
\frac{x}{s+1} \right)\frac{1}{s+1} \, ds,
\end{displaymath}
from which it immediately follows using (\ref{eqProb_GSD}) that 

\begin{displaymath}
 \phi_{\alpha}(x) \sim kx^{\alpha} \mbox{ as } x \rightarrow 0,
\end{displaymath}
for some constant $k$. This is the same small-size asymptotic
behaviour found in the fragmentation theory approach \cite{GLOM11,
  OGLM12}. We could not obtain the large-size asymptotics for
$\phi_{\alpha}(x)$ analytically. However, numerical analysis of the
solutions in Figure~\ref{i_0_1_2_3_IE_GSD} shows that they differ from
those obtained by the fragmentation equation approach that we believe
to be correct. The reason for this can be traced to the derivation of
the DFPE~(\ref{eqDFPE_GSD}), where not only do we adopt a MF
approach for nearest neighbour gap sizes, but we also neglect
longer-range correlations which are expected to be more prominent for
larger gaps created early in the growth process. An example of this effect
is, from Figure~\ref{1D_model}, is the creation of $g_3$ which arose from the 
nucleation $I_3$ and the fragmentation of gap of size ($g_1+g_3+g_5$).
This particular type of nucleation event is not included in the
DFPE~(\ref{eqDFPE_GSD}), which assumes that gaps arise from the fragmentation
of only two parents. Nevertheless, the results in Figure~\ref{i_0_1_2_3_IE_GSD}
 show that this approach captures much of the essential physics for the GSDs.

For the $i=0$ case, it is possible to use the fragmentation approach combined with 
Treat's results \cite{GLOM11,OGLM12,Treat} to obtain the GSD function

\begin{equation}
 \label{eqTreat_phi}
 \phi(x)=\frac{3x^2}{\Gamma(\frac{2}{3})\mu^3}\int_{(x/\mu)^3}^{\infty}u^{-4/3}e^{-u}\ du,
\end{equation} 
where 

\begin{displaymath}
 \mu = \frac{4}{3}\Gamma \left( \frac{2}{3} \right).
\end{displaymath}

In Figure~\ref{i_0_1_2_3_IE_GSD}, we also plot Treat's $\phi(z)$ from (\ref{eqTreat_phi})
to show how well our MF DFPE solutions work for $i=0$.

%Section -- The Non MF DFPE for the GSD

\section{NON MEAN-FIELD DFPE APPROACHES FOR THE GAP SIZE DISTRIBUTION\label{Sec4_nonMF_GSD}}

%Subsection 

\subsection{Unbiased DFPE}
\begin{figure}[H]
 \begin{center}
  \mbox{
        \begin{minipage}{1.3in}
        \scalebox{0.20}{\includegraphics{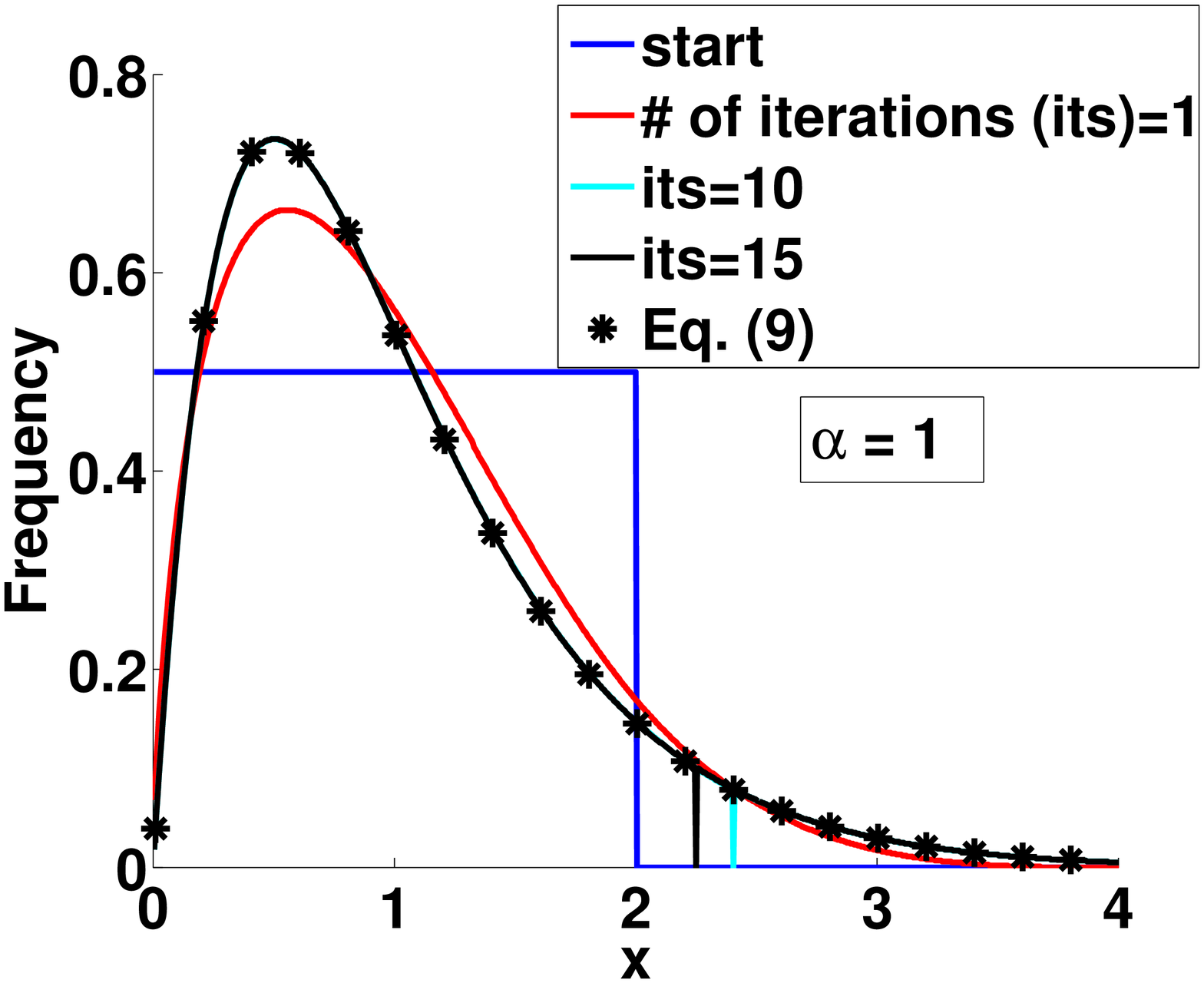}}\\
	    \end{minipage}
	    \qquad \qquad   %Note here, double qquad commands
	    \begin{minipage}{1.3in}
        \scalebox{0.20}{\includegraphics{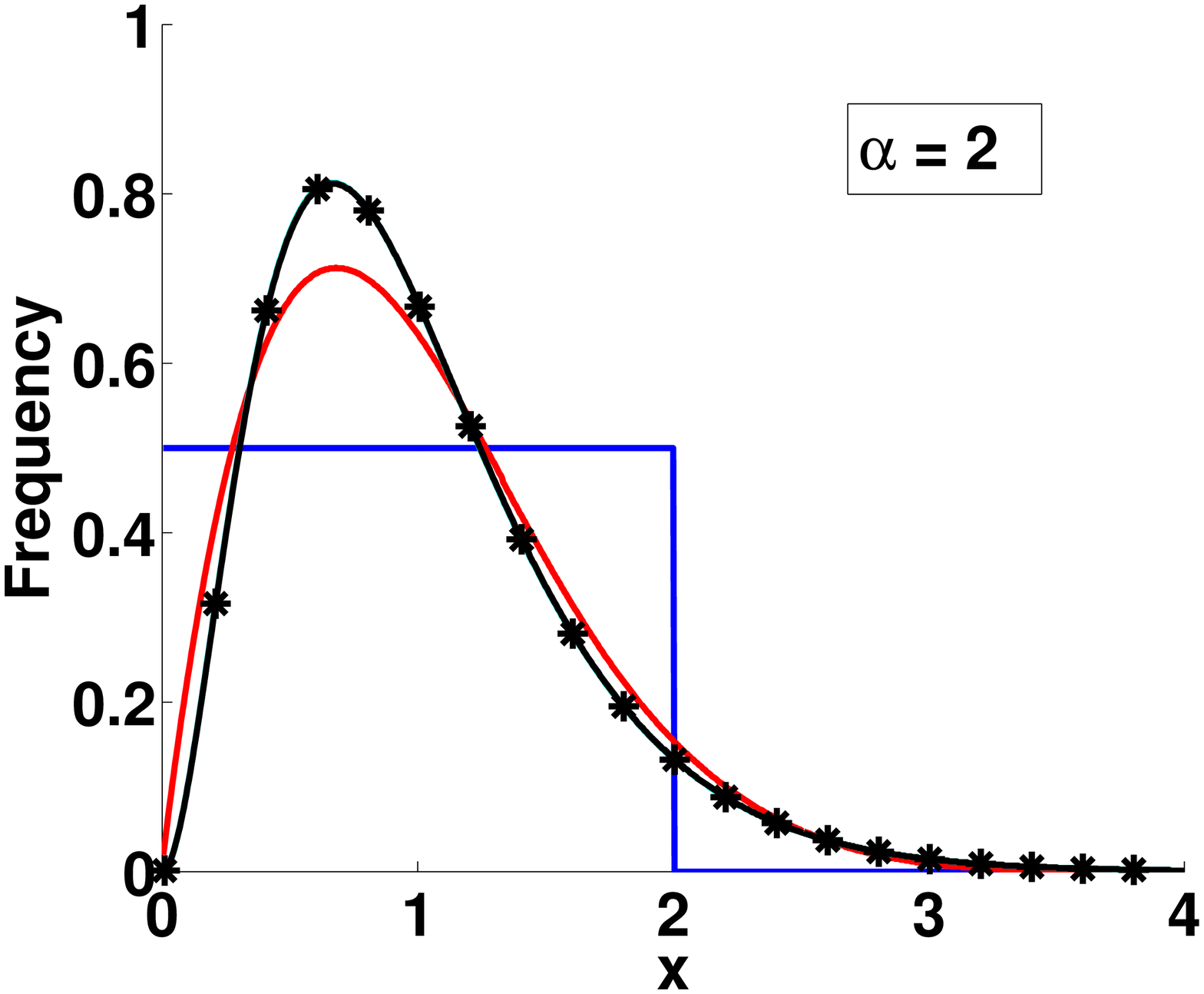}}\\
	    \end{minipage}
        }\\
  \end{center}
  \begin{center}
   \mbox{
        \begin{minipage}{1.3in}
        \scalebox{0.20}{\includegraphics{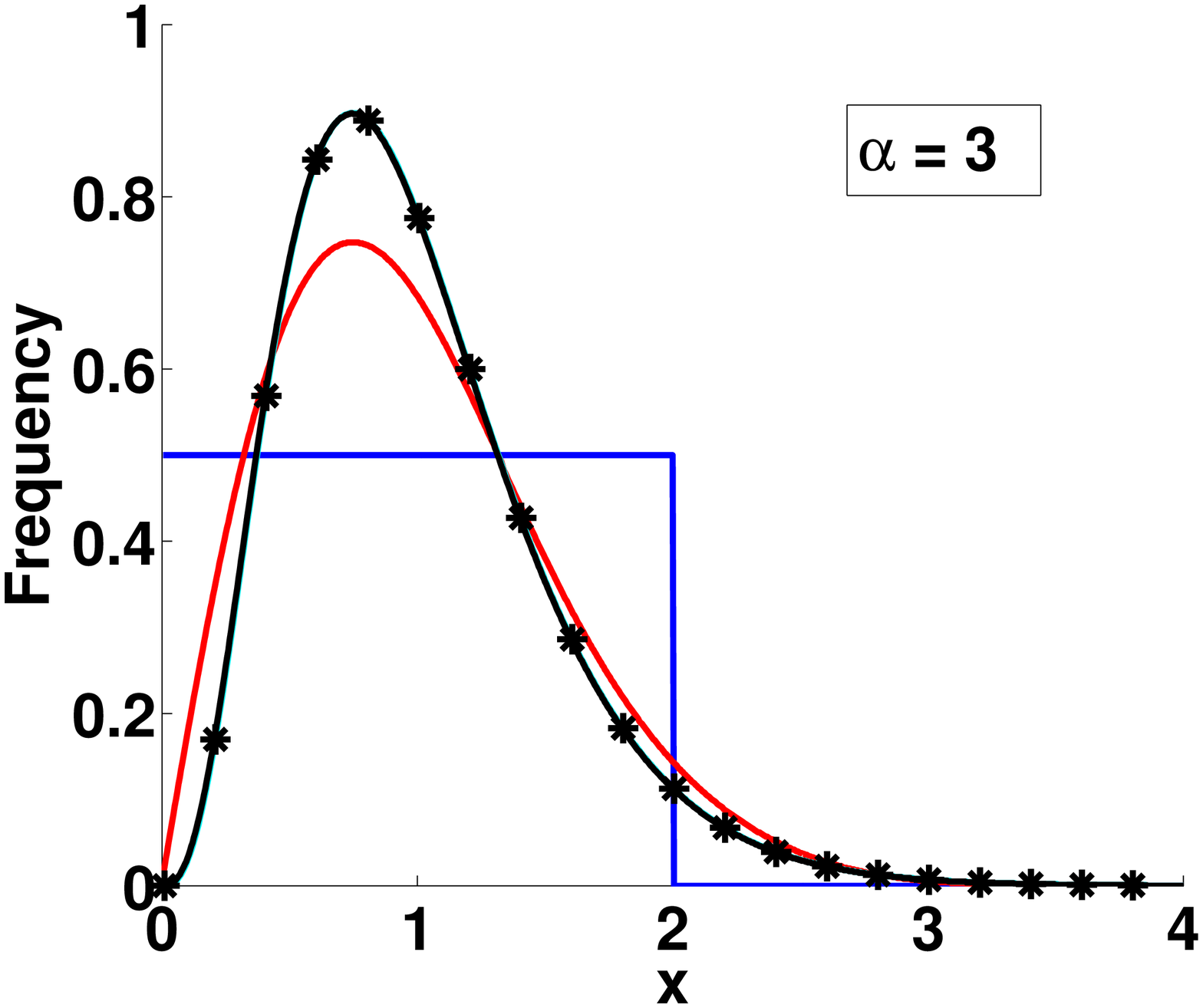}}\\
	    \end{minipage}
	    \qquad \qquad   %Note here, double qquad commands
	    \begin{minipage}{1.3in}
        \scalebox{0.20}{\includegraphics{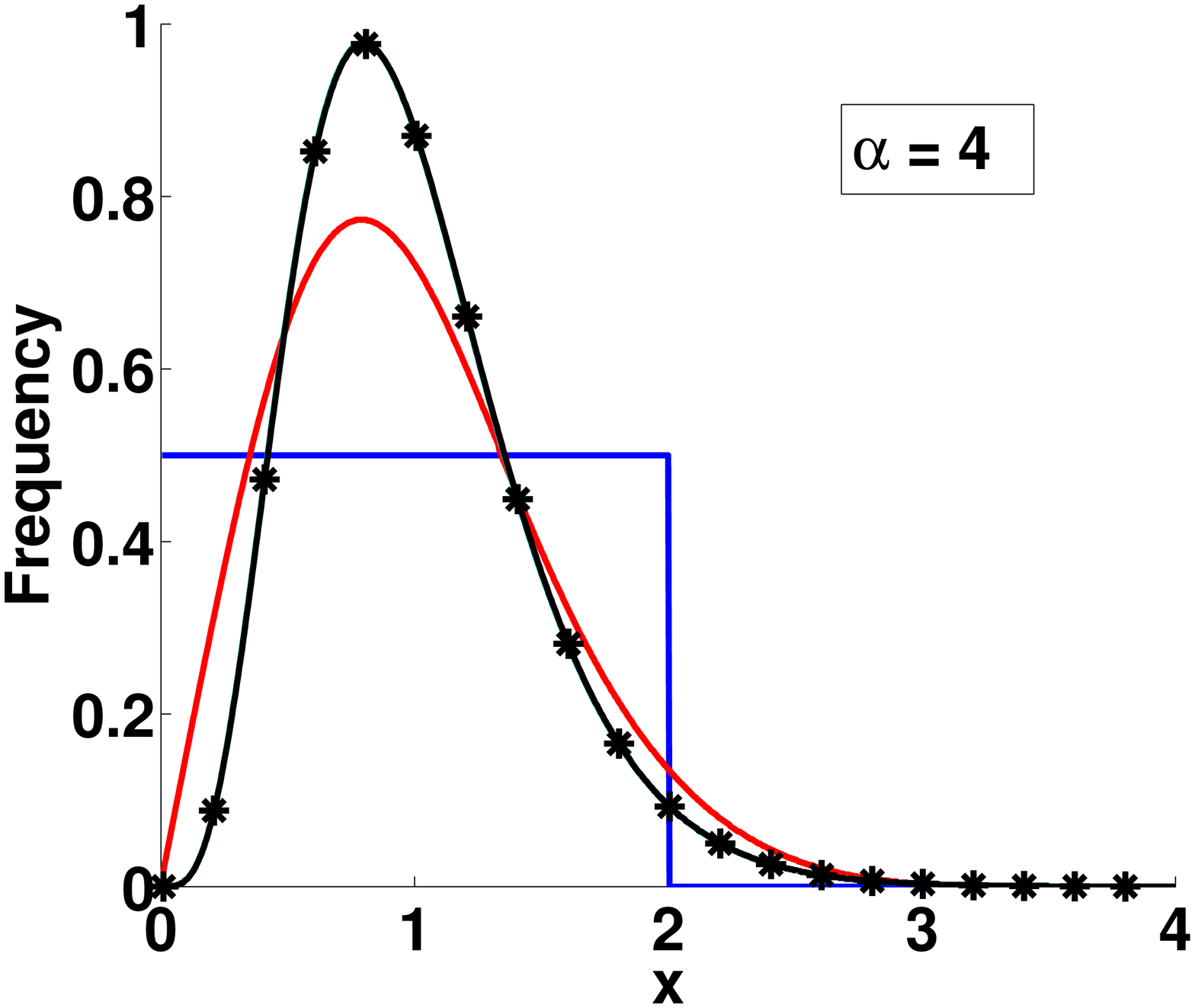}}\\
	    \end{minipage}
        }\\
 \end{center}
\caption{\label{alpha_all_iterations_nonMF} The evolution of gap size distribution under iteration 
of (\ref{eqIE_nonMF}) with $\alpha=1$, $2$, $3$, $4$.}
\end{figure}

In deriving equations~(\ref{eqDFPE_GSD}) and (\ref{eqIE_GSD}) in the previous section, we 
invoke a MF approximation for the size of the neighbouring gap, putting $y=1$. We could instead 
find the fixed point of the following DFPE that does not make a MF assumption:

\begin{equation}
  \label{eqDFPE_nonMF}
  x \eqd a(x_1+x_2),
\end{equation}
where the gaps $x_1$ and $x_2$ are, independently, drawn from the same distribution as $x$. 
As before $a$ is drawn from the probability distribution $f(a)$ of equation~(\ref{eqProb_GSD}). 
Then instead of (\ref{eqIE_GSD}) we have the following IE for the GSD $\phi(x)$:

 \begin{equation}
  \label{eqIE_nonMF}
   \phi(x) = \int_0^1 \int_0^{x/a} \phi \left( \frac{x}{a}-x_1 \right)\phi(x_1) \frac{f(a)}{a} \ dx_1 \ da,
 \end{equation} 
The derivation is similar to that for the IE~(\ref{eqIE_GSD}) shown in Appendix~\ref{App.1}.
The convergence of iterates of (\ref{eqIE_nonMF}) are shown in Figure~\ref{alpha_all_iterations_nonMF}.

Equation~(\ref{eqDFPE_nonMF}) with $f(a)$ as in (\ref{eqProb_GSD}) is considered by Dufresne \cite{Duf96}, 
where it is shown that the fixed point is given by a gamma distribution. Explicitly the fixed point 
probability distribution is

\begin{displaymath}
  \Gamma(\alpha+1,\nu,x)=\frac{x^{\alpha}\exp(-x/\nu)}{\Gamma(\alpha+1)\nu^{\alpha+1}}.
\end{displaymath}
The mean of the above gamma distribution is $(\alpha+1)\nu$ and so, by setting $\nu=1/(\alpha+1)$, we 
rescale $x$ to unity to obtain

\begin{equation}
 \label{eq_scaled_gamma_dist}
  \phi_{\alpha}(x)=\frac{(\alpha+1)^{\alpha+1}}{\Gamma(\alpha+1)}x^{\alpha}e^{-(\alpha+1)x}.
\end{equation}
Note that if we assume (\ref{eq_scaled_gamma_dist}), then for small $x$ we obtain $\phi(x) \sim kx^{\alpha}$ 
for some constant $k$. In Figure~\ref{alpha_all_iterations_nonMF}, this gamma distribution is shown by the 
stars. It is apparent that the iterations converge to the form (\ref{eq_scaled_gamma_dist}) and, as we see, 
it is not surprising to confirm the result obtained by Dufresne \cite{Duf96}.

%Subsection 

\subsection{Fragmentation bias for the non-MF DFPE}

The IE model presented in Eqn.~({\ref{eqIE_nonMF}) for the GSD is not appropriate 
for the island nucleation process, since we know from MC simulations \cite{OGLM12} that larger gaps are 
fragmented by nucleation events more often than smaller ones. 
We account for this effect in the following way. Referring to the non-MF DFPE Eqn.~({\ref{eqDFPE_nonMF}), 
we still wish to draw $x_1$ from $\phi(x_1)$ in an unbiased way. However, $x_2$ is not unaffected by the
value of $x_1$, and should not be drawn simply from $\phi(x_2)$ as we did in Eqn.~({\ref{eqIE_nonMF}).
Instead we draw it from a skewed distribution $(x_1+x_2)^{2\alpha+1}.\phi(x_2)$, reflecting the fact that a parent 
gap of size $x_1+x_2$ is fragmented with probability $(x_1+x_2)^{2\alpha+1}$ from the integration of monomer density 
raised to the power $\alpha$ in the gap \cite{OGLM12, GLOM11, BM96}. With this,
we derive the following, correctly biased non-MF IE:

\begin{align}
 \label{eqIE_nonMF_bias}
  \phi(x) & = \int_0^1 \int_0^{x/a} \phi\left( \frac{x}{a}- x_1 \right) \phi(x_1)f(a) \nonumber \\
          & \times \frac{x^{2\alpha+1}}{a^{2\alpha+2}} \ da \ dx_1.
\end{align}

We now compare the fixed point probability distribution of equation~(\ref{eqIE_nonMF_bias}) with those 
from the MF approximation~(\ref{eqIE_GSD}) and from (\ref{eqIE_nonMF}) above; 
see Figure~\ref{alpha_all_comparisons}. Note that the effect of the bias is to skew the distribution 
away from that of equation~(\ref{eqIE_nonMF}), which over-represents small gaps, towards that of the 
MF approximation (see also Figure~\ref{i_0_1_2_3_IE_GSD}). 
The reason for this can be found in the biased form $(x_1+x_2)^{2\alpha+1}.\phi(x_2)$ which becomes more sharply peaked
for larger $\alpha$, so its replacement by a single value in the MF equations is increasingly justified as $\alpha$ is increased.
This behaviour is apparent in the results shown in Figure~\ref{alpha_all_comparisons}, and indeed even for $\alpha=1$ 
(corresponding to $i=0$) it is a very good approximation. Therefore this fragmentation bias vindicates the use of the MF
approximation for the GSD, which allows us to proceed with some confidence to consider the CZD from the same MF perspective.

\begin{figure}[H]
 \begin{center}
  \mbox{
        \begin{minipage}{1.3in}
        \scalebox{0.20}{\includegraphics{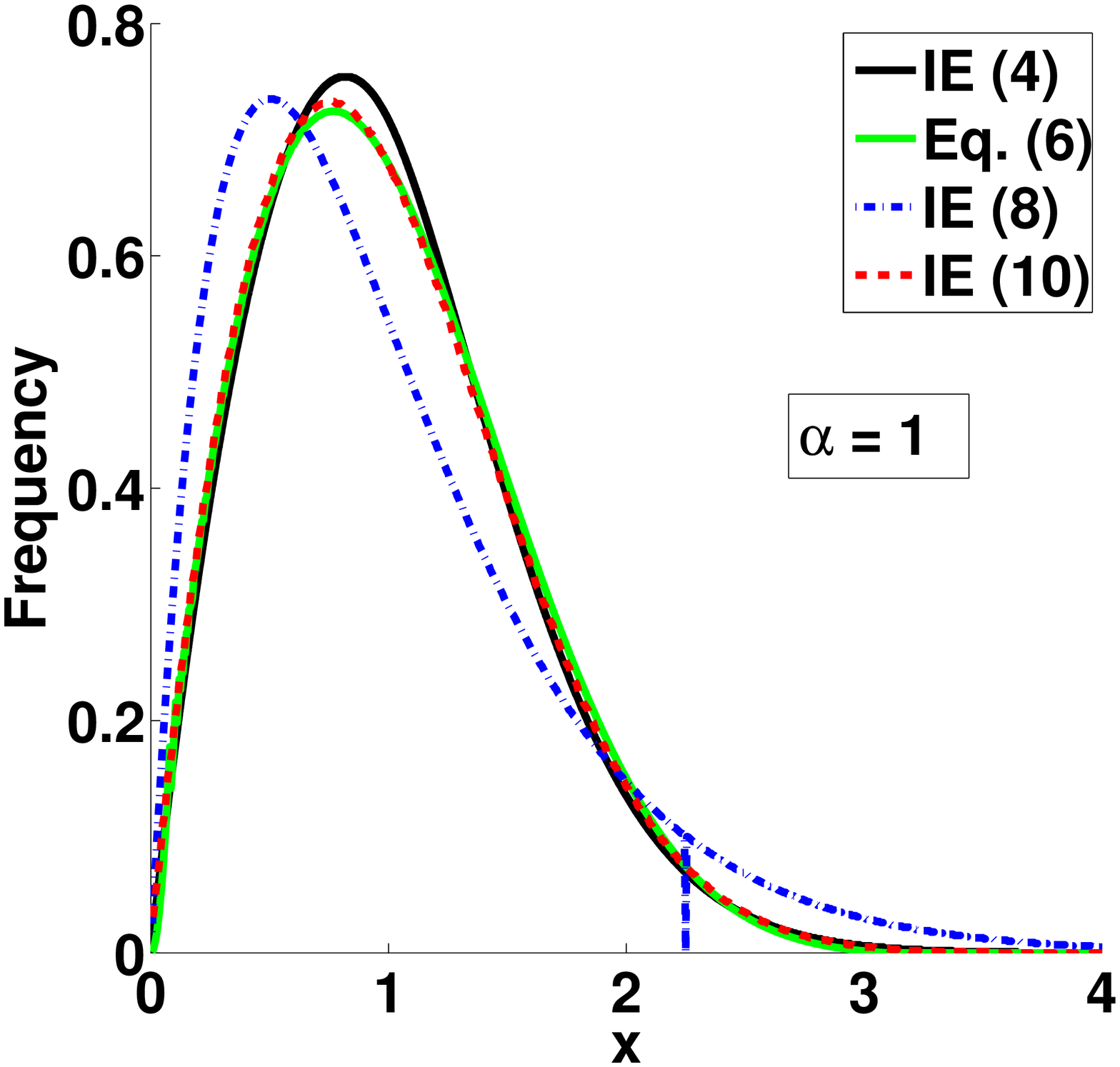}}\\
	    \end{minipage}
	    \qquad \qquad   %Note here, double qquad commands
	    \begin{minipage}{1.3in}
        \scalebox{0.20}{\includegraphics{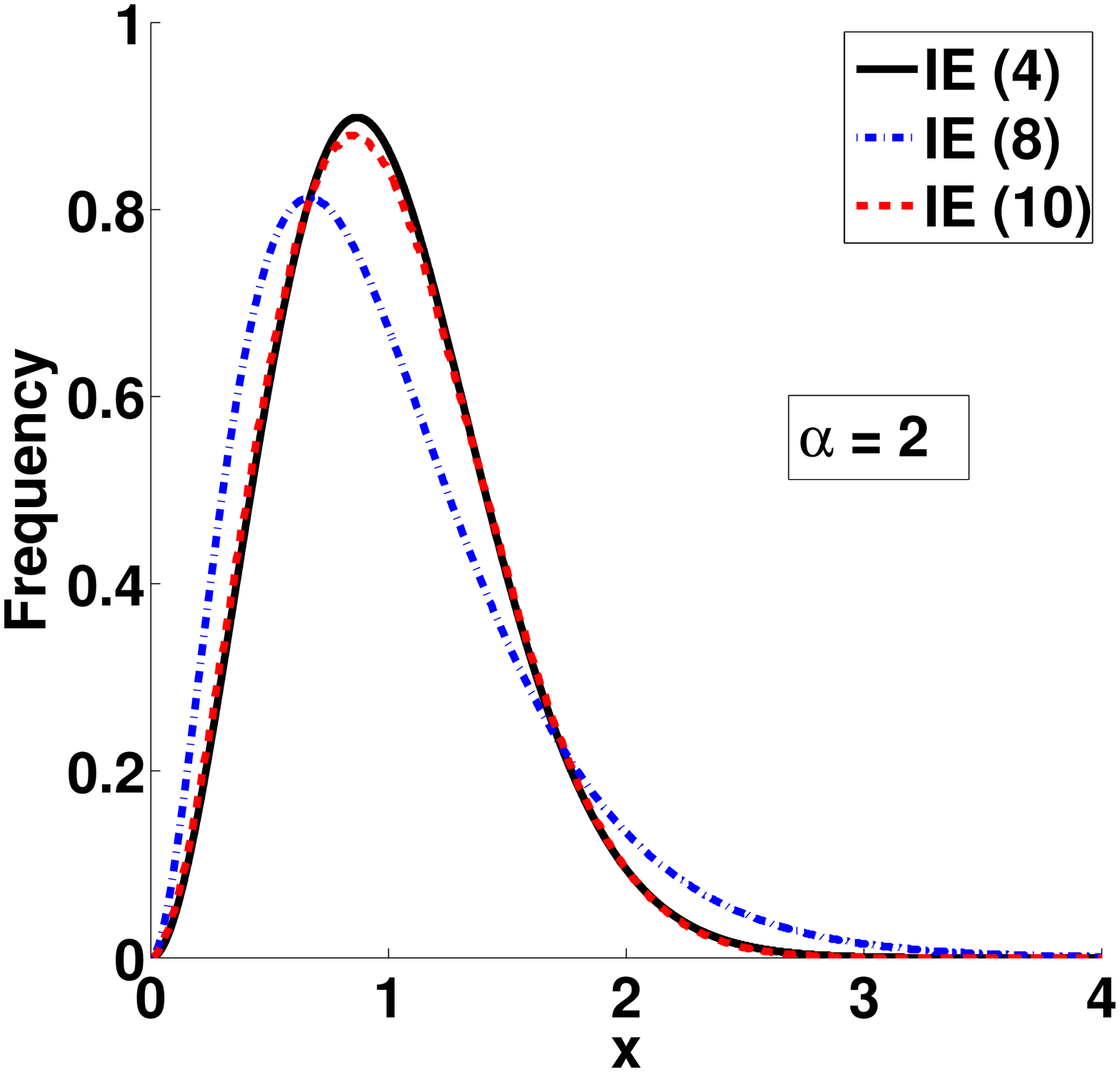}}\\
	    \end{minipage}
        }\\
  \end{center}
  \begin{center}
   \mbox{
        \begin{minipage}{1.3in}
        \scalebox{0.20}{\includegraphics{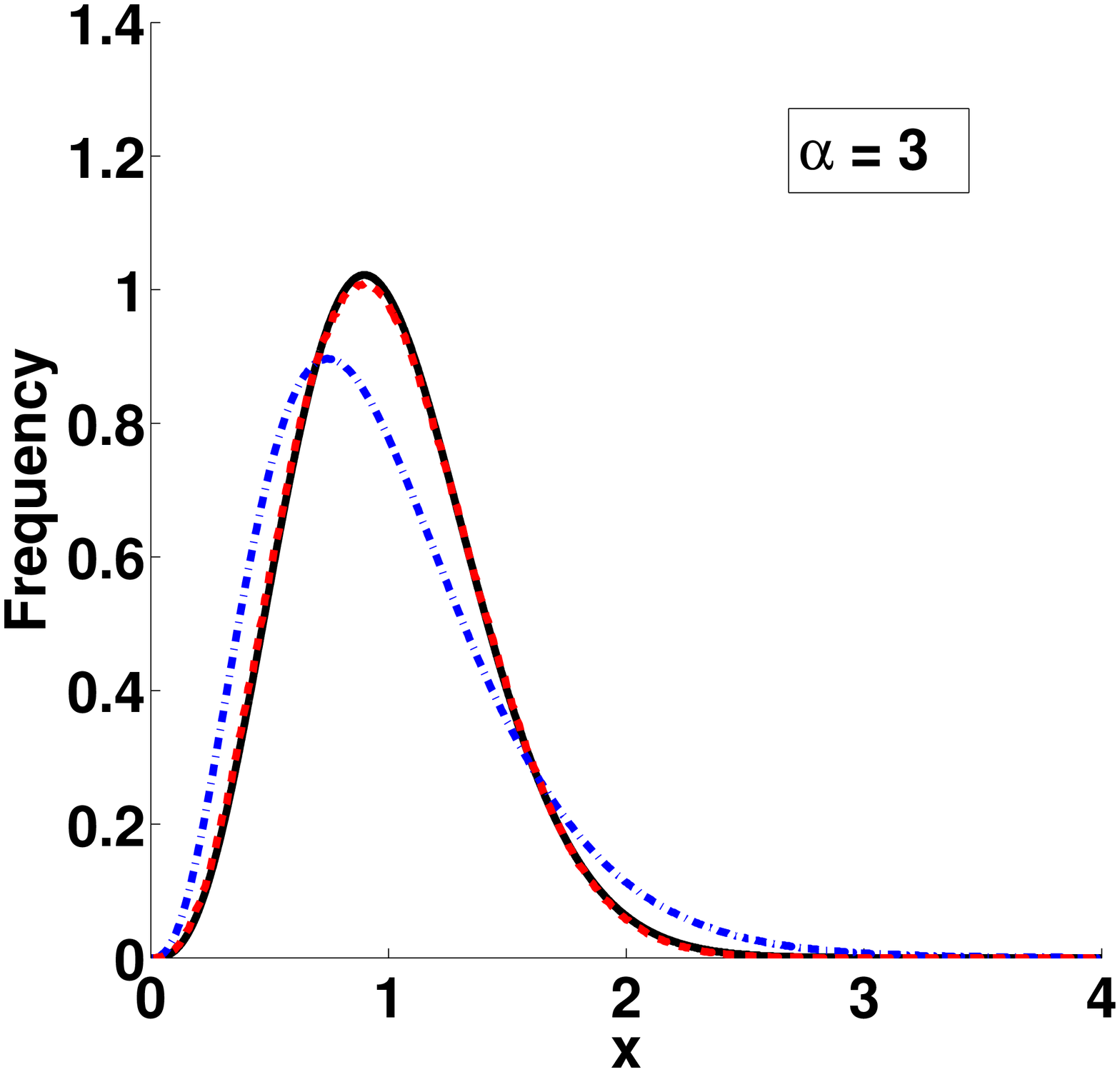}}\\
	    \end{minipage}
	    \qquad \qquad   %Note here, double qquad commands
	    \begin{minipage}{1.3in}
        \scalebox{0.20}{\includegraphics{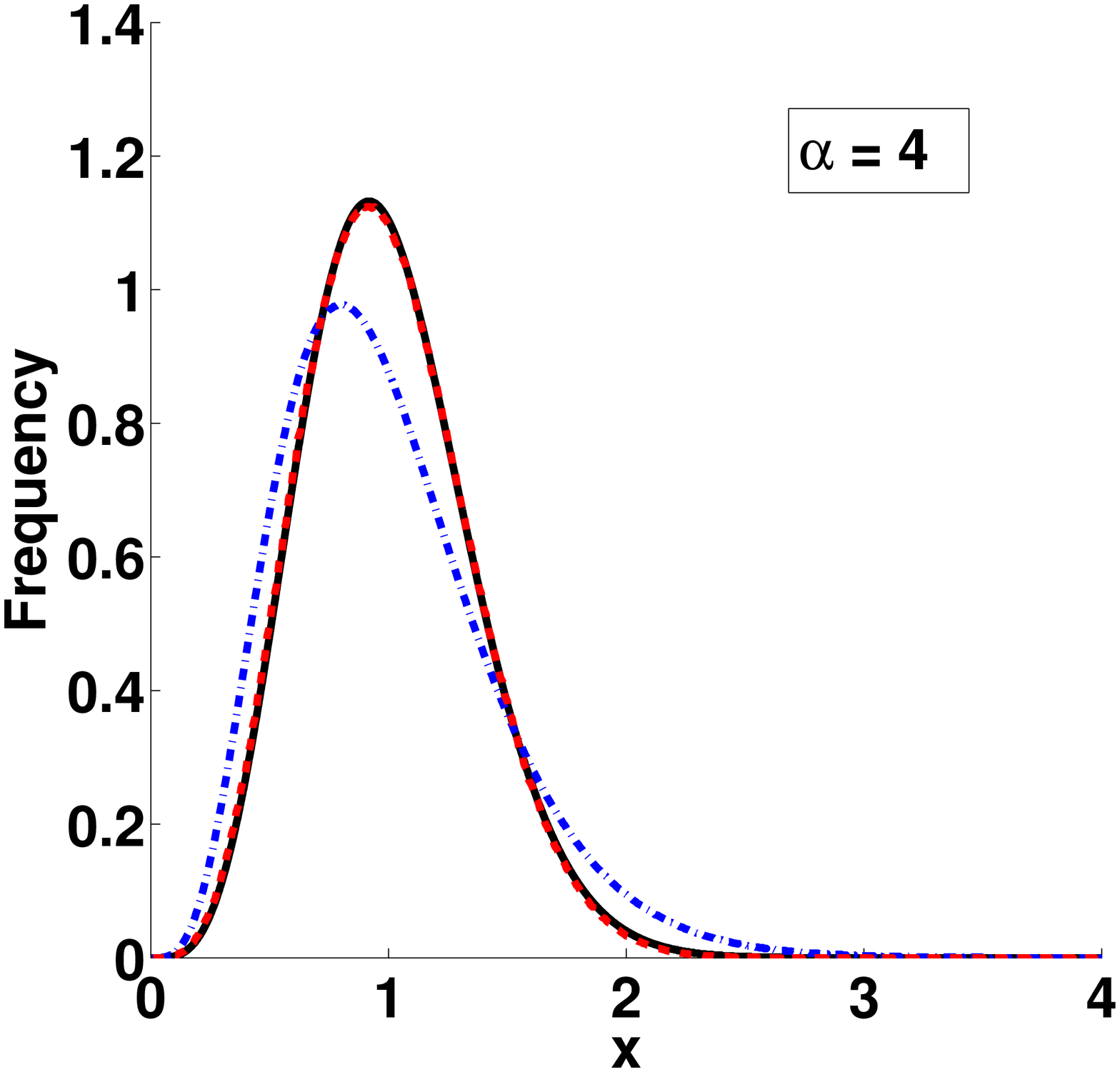}}\\
	    \end{minipage}
        }\\
 \end{center}
\caption{\label{alpha_all_comparisons} Comparison of the integral equations~(\ref{eqIE_GSD}), 
(\ref{eqIE_nonMF}) and (\ref{eqIE_nonMF_bias}) and, for the $\alpha=1$ ($i=0$) case, Treat's 
solution~(\ref{eqTreat_phi}) for the various gap size models with $\alpha=1$, $2$, $3$, $4$.}
\end{figure}

%Section -- The MF DFPE for the CZD

\section{THE MEAN-FIELD DFPE FOR THE CAPTURE ZONE DISTRIBUTION\label{Sec5_MF_CZD}}
We turn now to consider the evolution of the capture zones in the
system.  One possible approach is to assume that neighbouring gaps are
not correlated in size, so that the capture zone distributions (CZDs)
can be calculated from convolutions of the related GSDs
\cite{BM96,GLOM11,OGLM12}. However, here we prefer to progress in the
same spirit as above, and use the MF approach to construct a
DFPE for the capture zones, since this approach might be transferable
to higher dimension substrates \cite{comment}.  Referring back to
Figure~\ref{1D_model}, we see that the capture zone $C_5$ was created
by the nucleation of island $I_5$. Prior to this, the zones $C_3$ and
$C_4$ were larger, so that the creation of $C_5$ can be viewed as the
fragmentation of part of $C_3$ (the part to the right of island $I_3$)
and part of $C_4$ (to the left of $I_4$). In general we do not know
how much of the neighbouring capture zones to take, nor indeed how
large these zones are. However, we can again invoke a MF
approximation for these nearest neighbour correlations to find the
following DFPE for a general capture zone $c$:

\begin{equation}
 \label{eqDFPE_CZD}
 c \eqd \frac{1}{2}(a_1+a_2)(1+c).
\end{equation}
The proportions $a_1$ and $a_2$ are independently drawn from $f(a)$ of equation~(\ref{eqProb_GSD}). 
An equivalent IE, like that of equation~(\ref{eqIE_GSD}), can readily be identified for 
equation~(\ref{eqDFPE_CZD}).

In Figure~\ref{i_0_1_2_3_IE_CZD} we compare the CZDs obtained as fixed
points of equation~(\ref{eqDFPE_CZD}) with those from the MC
simulations \cite{OGLM12}. Again we find excellent agreement,
particularly for $i=0$ and $i=1$. We also plot the GWS~(\ref{GWS}) as a
convenient analytical form. We see that the solution of equation
(\ref{eqDFPE_CZD}) fits the data at least as well as, and in the case
of $i=0$ much better than, the GWS. Whilst the validity of the GWS has
been questioned as mentioned above in Section I, it is a useful benchmark for
comparisons to MC data \cite{GPE11, OR11}.

\begin{figure}[h!]
 \begin{center}
  \mbox{
        \begin{minipage}{1.3in}
        \scalebox{0.20}{\includegraphics{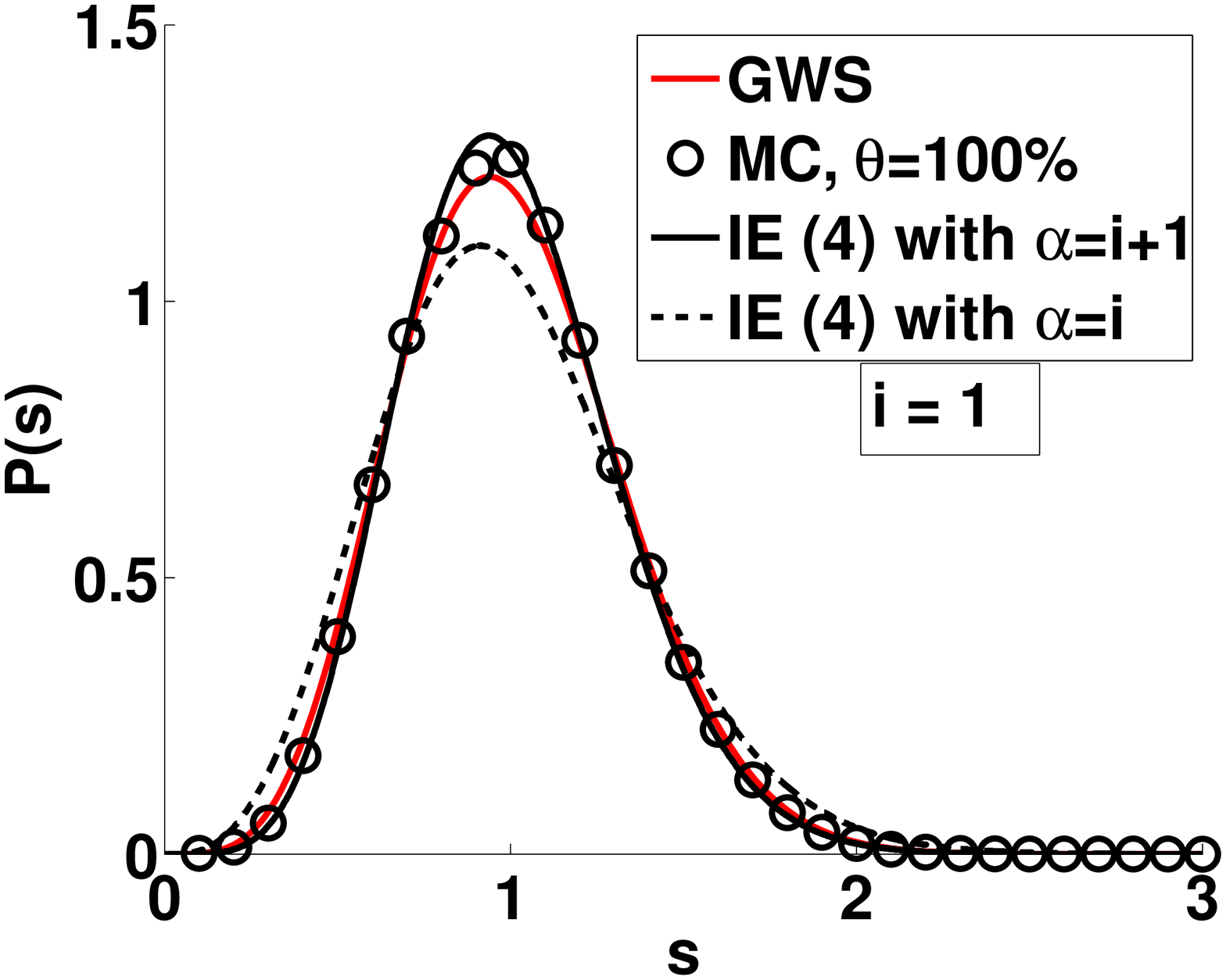}}\\
	    \end{minipage}
	    \qquad \qquad   %Note here, double qquad commands
	    \begin{minipage}{1.3in}
        \scalebox{0.20}{\includegraphics{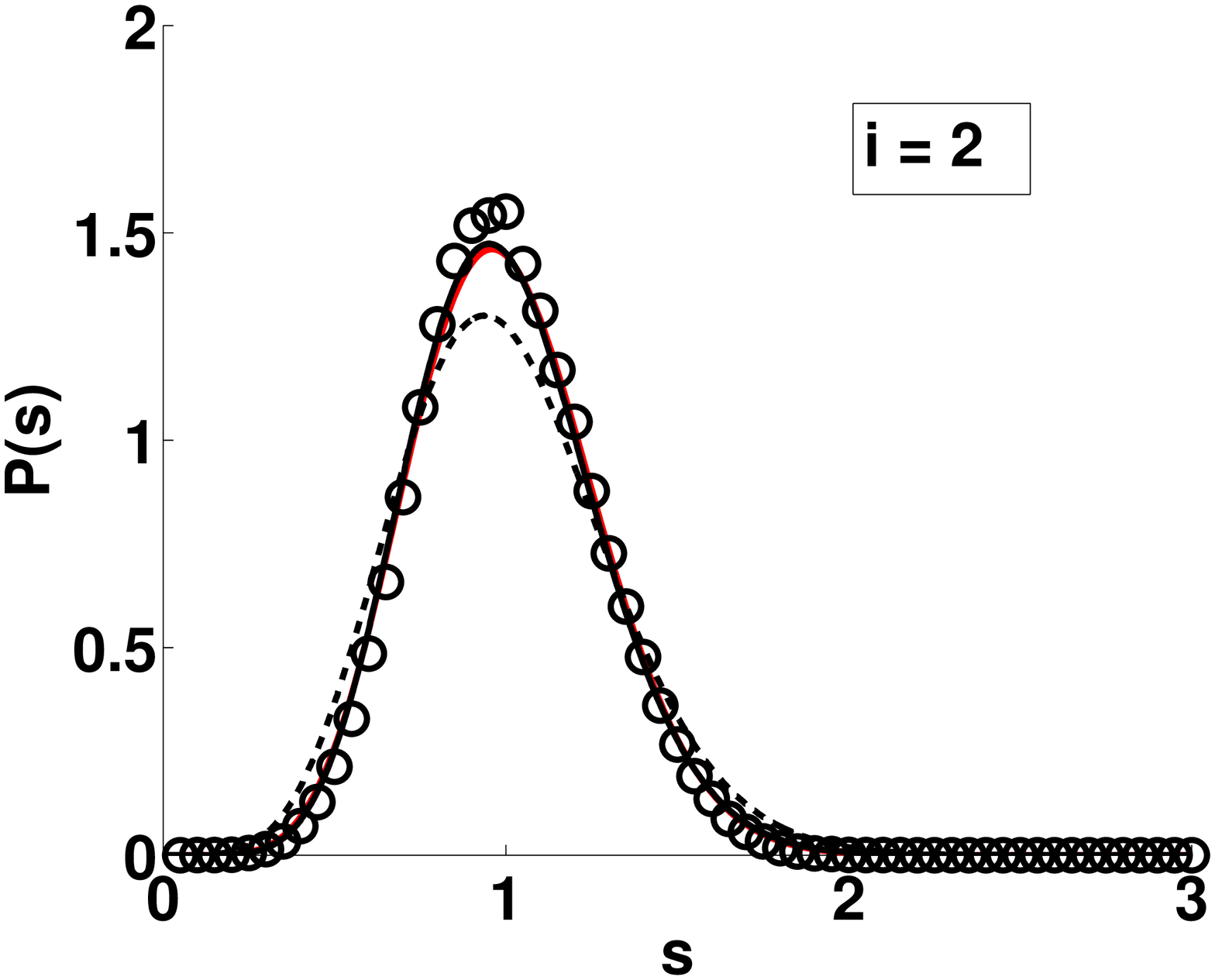}}\\
	    \end{minipage}
        }\\
  \end{center}
  \begin{center}
   \mbox{
        \begin{minipage}{1.3in}
        \scalebox{0.20}{\includegraphics{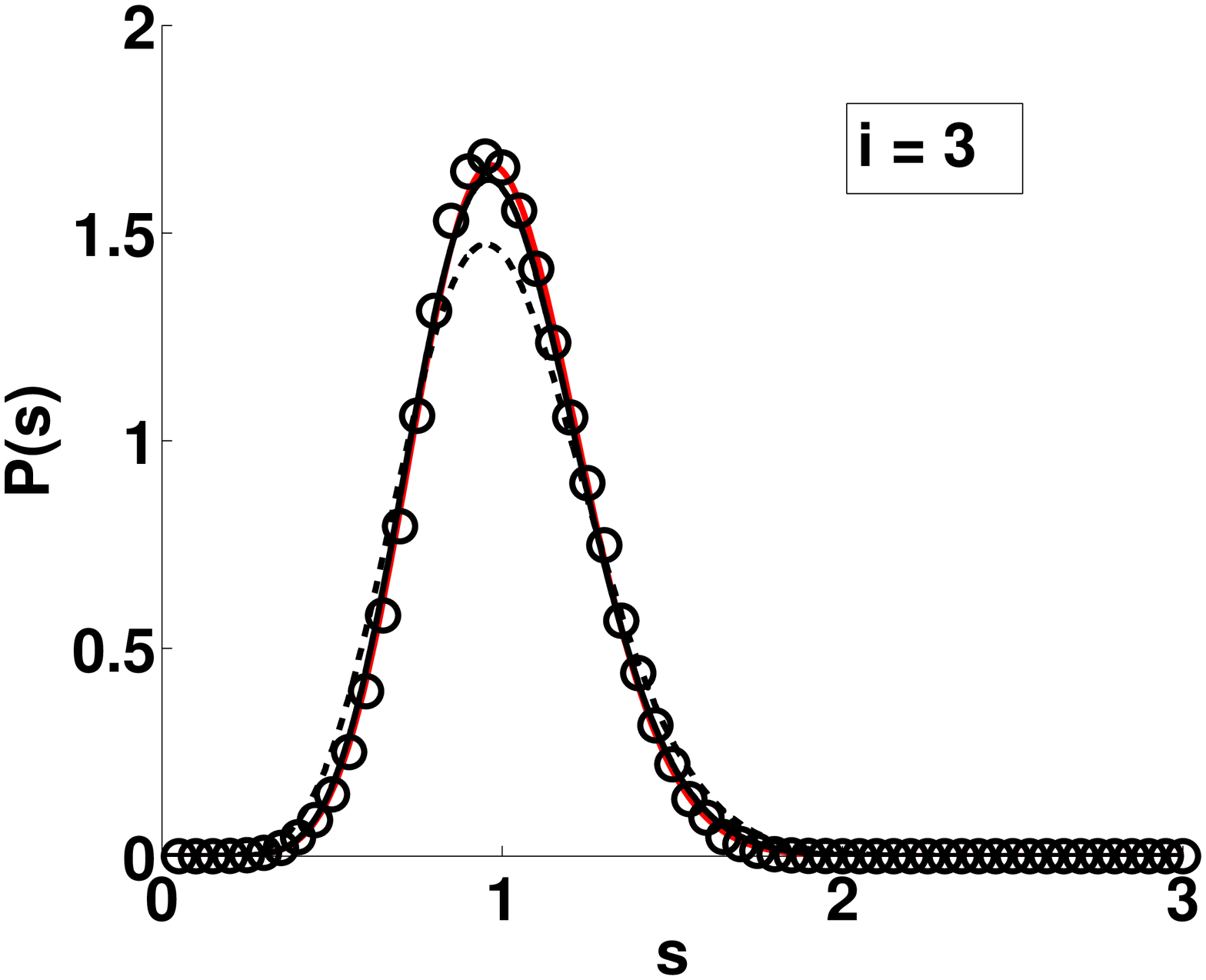}}\\
	    \end{minipage}
	    \qquad \qquad   %Note here, double qquad commands
	    \begin{minipage}{1.3in}
        \scalebox{0.20}{\includegraphics{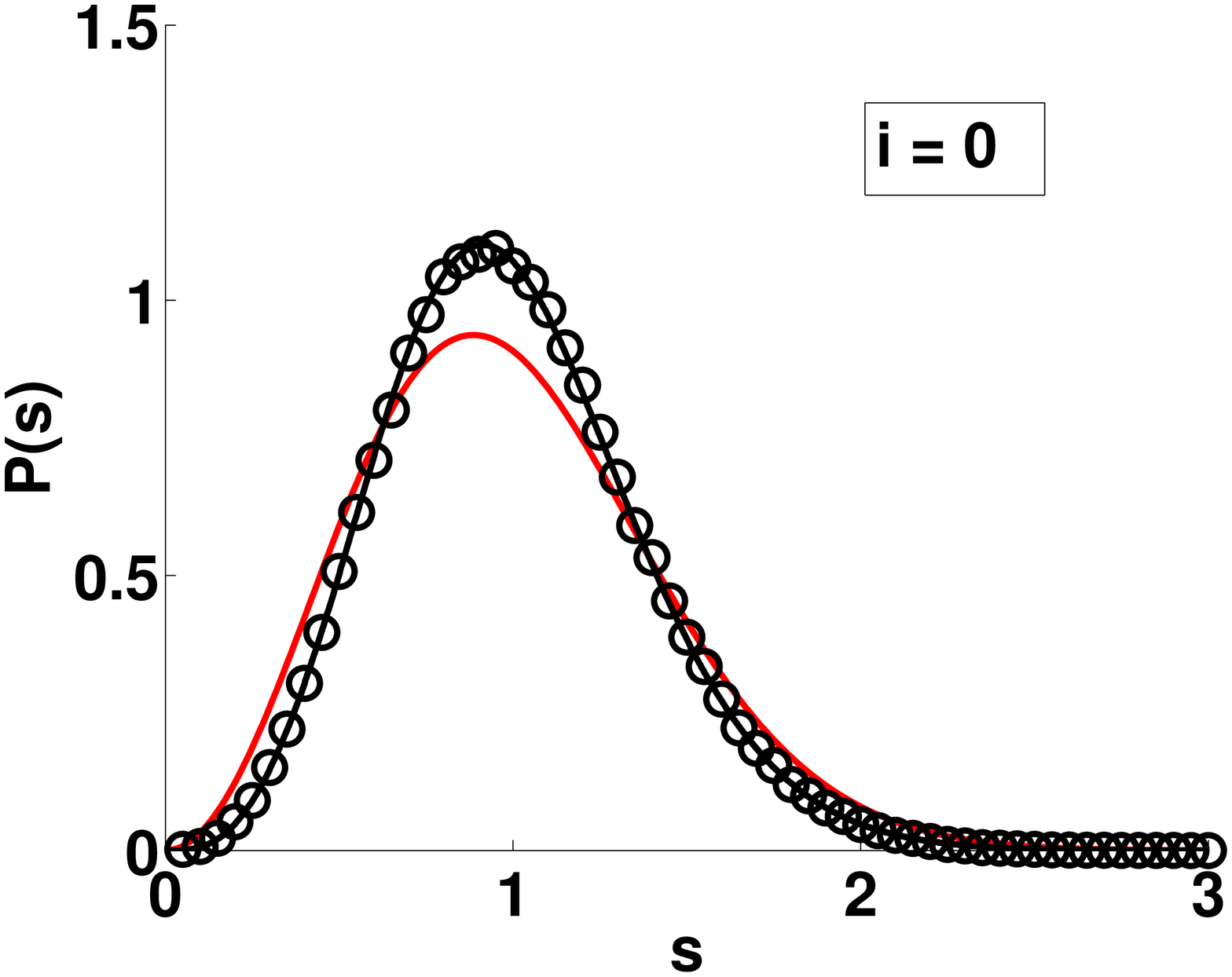}}\\
	    \end{minipage}
        }\\
 \end{center}
\caption{\label{i_0_1_2_3_IE_CZD}The CZDs compared to histograms of MC data \cite{OGLM12} for various 
critical island size $i$, taken at nominal coverage $\theta=100$\%. The solid curves are the solutions 
to equation~(\ref{eqDFPE_CZD}) for $\alpha=i+1$ with $i=0,1,2,3$, and the broken lines are for 
$\alpha=i$ with $i=1,2,3$.}
\end{figure}

We can easily quantify the performance of the solutions using the
moments $S_m$ of the CZDs. Typically, when the GSDs, CZDs and/or ISDs are measured,
the interest is the shape of the distribution and whether there is good data collapse to scaling forms 
which reveal nucleation and growth mechanisms. A distribution can be identified by a number of features 
such as the mean (first moment), the variance (second moment), the skewness 
(third moment) and the kurtosis (fourth moment) etc. In other words, the moments of 
a distribution can help to characterise its nature even if the full distribution is
unknown, for example in a limited set of experimental data. 
Following [\onlinecite{LJC10}], from equation~(\ref{eqDFPE_CZD}) 
we find the following recursive relationship:

\begin{equation}
 \label{eqDFPE_CZD_moments}
 S_m = \left(\frac{1}{2}\right)^m \sum_{k=0}^m \frac{m!}{k!(m-k)!} B_{m-k}B_k \sum_{p=0}^m \frac{m!}{p!(m-p)!}S_p,
\end{equation}
where

\begin{displaymath}
 B_m = \frac{B(m+\alpha+1,\alpha+1)}{B(\alpha+1,\alpha+1)}, 
\end{displaymath}
and $B(m,n)$ is the Beta function as in equation~(\ref{eqProb_GSD}).

The moments of the GWS, Eqn.~(\ref{GWS}), are given by

\begin{equation}
 \label{eqGWS_moments}
 G_m = \frac{ \Gamma(i+3/2)^{m-1} \Gamma(i+(m+3)/2)}{(i+1)!^m}.
\end{equation}

In Table~\ref{i_0_1_2_3_moments_CZD} we compare the moments calculated from equations~(\ref{eqDFPE_CZD_moments}) 
and (\ref{eqGWS_moments}) alongside those taken from our MC simulations \cite{OGLM12} for $i=0,1,2,3$. 
These confirm the superiority of the DFPEs, notably for $\alpha = i+1$ implying a greater significance for 
nucleation driven by monomer diffusion.

% Moments for i = 0-3 CZD
\begin{table}[!h]
 \begin{center}
  \begin{tabular}{c|cccc}
   $m$ & $G_m$ & $S_m$ \footnote{$\alpha_n=i$} & $S_m$ \footnote{$\alpha_n=i+1$} & MC \footnote{Point islands, $\theta=100$\%} \\
   \hline
   \underline{$i=0$} & & & & \\
   $2$ & $1.178$ & - & $1.138$ & $1.134 \pm 0.001$ \\
   $3$ & $1.571$ & - & $1.439$ & $1.425 \pm 0.001$ \\
   $4$ & $2.313$ & - & $1.989$ & $1.949 \pm 0.001$ \\
   \hline
   \underline{$i=1$} & & & & \\
   $2$ & $1.105$ & $1.138$ & $1.098$ & $1.098 \pm 0.001$ \\
   $3$ & $1.325$ & $1.439$ & $1.305$ & $1.307 \pm 0.001$ \\
   $4$ & $1.708$ & $1.989$ & $1.665$ & $1.666 \pm 0.001$ \\
   \hline
   \underline{$i=2$} & & & & \\
   $2$ & $1.074$ & $1.098$ & $1.076$ & $1.066 \pm 0.001$ \\
   $3$ & $1.227$ & $1.305$ & $1.234$ & $1.202 \pm 0.001$ \\
   $4$ & $1.483$ & $1.665$ & $1.500$ & $1.425 \pm 0.001$ \\
   \hline
   \underline{$i=3$} & & & & \\
   $2$ & $1.057$ & $1.076$ & $1.062$ & $1.056 \pm 0.001$ \\
   $3$ & $1.175$ & $1.234$ & $1.190$ & $1.169 \pm 0.001$ \\
   $4$ & $1.366$ & $1.500$ & $1.401$ & $1.352 \pm 0.001$ \\
  \end{tabular}
   \caption{Moments of the CZDs for $i=0$, $1$, $2$ and $3$ from the DFPE (\ref{eqDFPE_CZD_moments}) with 
$\alpha_n=i+1$ or $\alpha_n=i$ (if appropriate), and from the MC simulations taken at 
$\theta=100$\%.}\label{i_0_1_2_3_moments_CZD}
 
 \end{center}
\end{table}

%Section -- Discussion and Conclusions

\section{SUMMARY AND CONCLUSIONS\label{Sec6_summary}}

In summary, we have presented distributional fixed point equations (DFPEs)
and their equivalent, integral equations (IEs) for the nucleation of point 
islands in one dimension. The approach develops a new retrospective view 
of how the inter-island gaps and capture zones have developed from the 
fragmentation of larger entities. 

To help validate our approach, we
have carried out Monte Carlo (MC) simulations of the one-dimensional (1-D)
point-island model for island nucleation and growth for $i=0,1,2,3$ and values of $R$
ranging from $10^7$ to $10^9$. We find that with higher values of $R$ there is good scale-invariance
for the gap size distributions (GSDs) and capture zone distributions (CZDs) at coverages 
($\theta \le 20$\%), where the scaling form depends on critical island size $i$. Significantly, we also
find little to distinguish these distributions from those of simulations with more realistic extended islands,
so that whilst our subsequent DFPE development is focused on point islands, it can equally apply to the extended 
island case, at least for low substrate coverage.

We first developed a mean field (MF) approach to the nucleation of gaps between islands on the substrate,
arguing for the simple DFPE of Eqn.~(\ref{eqDFPE_GSD}) and its associate IE of Eqn.~(\ref{eqIE_GSD}). We found good
comparisons between the converged solutions and the MC data, suggesting that the model has a reasonable
physical basis. Exploring the idea further, we next considered the non-MF DFPE of Eqn.~(\ref{eqDFPE_nonMF}). 
Without using any selection bias for the parents we showed that solutions to the associated IE Eqn.~(\ref{eqIE_nonMF})
are gamma distributions \cite{Duf96}. However, including the bias towards fragmenting large parents arising from the
fragmentation probability $f(a)$ in Eqn.~(\ref{eqProb_GSD}), we found that the solutions are drawn back to those of the initial
MF version, justifying this approximation. Interestingly, even for $i=0$ we found that the MF solution works well and is
reasonably close to the exact fragmentation theory solution for this case \cite{Treat}.

We also considered the DFPE~(\ref{eqDFPE_CZD}) and IE in the form of 
(\ref{eqIE_GSD}) for the CZD, following in the same MF spirit. 
This allowed a closed form for the CZD to be developed, unlike previous approaches \cite{BM96,GLOM11,OGLM12,GPE11}
where the CZD is explicitly derived from convolution of the GSD. The solutions 
to our equations compare well to MC simulation data, performing at least as
well as the Generalised Wigner Surmise (GWS) \cite{PE07}, and notably better for the case of $i=0$. 
The recursive form of DFPE also allowed calculation of moments of the distributions, values which might be
useful in the future for assessing limited experimental data.

Our presentation of DFPEs and associated IEs for island nucleation in 1-D provides a fresh
perspective on why scaling emerges in this non-equilibrium growth system. Furthermore, we hope that a similar
approach might be possible in higher dimensions too. Whilst the utility of inter-island gaps is perhaps limited to the 1-D
case, capture zones still underpin island growth rates in higher dimensions \cite{BM96, MR00, EB02, Li10}, so that it might be possible in future work to 
find an equivalent closed-form DFPE for the CZD on higher dimensional substrates.

%Appendices

\appendix

\section{Derivation of the integral equation~(\ref{eqIE_GSD})\label{App.1}}

Following the analysis \cite{PW,Seba07}, we obtain an IE which is equivalent to the DFPE~(\ref{eqDFPE_GSD})

\begin{proposition}
 For the gap size distribution, $\phi(x)$, the following integral equation
 
 \begin{displaymath}
  \phi(x) = \int_0^{\min(x,1)} \phi \left( \frac{x}{a}-1 \right) \frac{f(a)}{a} da,
 \end{displaymath} 
is derived from the distributional fixed point equation~(\ref{eqDFPE_GSD}).
\end{proposition}

We give a proof of this proposition. Let $\Phi$ be the cumulative distribution function (CDF) that corresponds to the density function $\phi$.
Then $\Phi(x)=0$ for all $x \le 0$ and we have
 \begin{align*}
\Phi(x)
 & = Prob[x_1 \leq x] \\
 & = Prob[a(1+x_1) \leq x]  \\
 & = E[ \ Prob[a(1+x_1) \leq x \ | \ a] \ ]  \\
 & = \int_0^1 Prob[a(1+x_1) \leq x] f(a) da  \\
 & = \int_0^1 Prob[x_1 \leq x/a -1] f(a) da \\
 & = \int_0^1 \Phi(x/a -1)H(x/a -1) f(a) da, 
\end{align*}
where $H(\cdot)$ is the Heaviside function since $\Phi(x_1)=Prob[x_1 \le a]$ and since $\Phi(x)=0$ if $x \le 0$. 
Hence the CDF satisfies

\begin{equation}
 \label{eq_IE_CDF}
 \Phi(x) = \int_0^{\min(x,1)} \Phi \left( \frac{x}{a} - 1 \right) f(a)da.
\end{equation}

The change variables to $w=x/a-1$ yields

\begin{displaymath}
 \Phi(x) = x \int_{\max(0,x-1)}^{\infty} \Phi(w) f\left( \frac{x}{w+1} \right) \frac{1}{(w+1)^2} \ dw.
\end{displaymath}
Given the form~(\ref{eqProb_GSD}), the differentiability of $f$ means that we can use the Leibniz rule to
establish that $\Phi'(x)=\phi(x)$ exists for each of the cases $x>1$ and $x<1$. Returning to 
equation~(\ref{eq_IE_CDF}), we obtain, for $x < 1$,

\begin{align*}
 \phi(x) = \Phi'(x) & = \Phi(0)f(x) + \int_0^x \Phi'\left( \frac{x}{a}-1 \right) \frac{f(a)}{a} \ da \\
                    & = \int_0^x \phi\left( \frac{x}{a}-1 \right) \frac{f(a)}{a} \ da,
\end{align*}
and, for $x>1$,

\begin{displaymath}
 \phi(x)=\Phi'(x) = \int_0^1 \phi\left( \frac{x}{a}-1 \right) \frac{f(a)}{a} \ da.
\end{displaymath}

Taking left-sided and right-sided limits, we deduce that

\begin{displaymath}
\phi(1)= \lim_{x \rightarrow 1} \phi(x) = \int_0^1 \phi\left( \frac{1}{a}-1 \right) \frac{f(a)}{a} \ da,
\end{displaymath}
and the stated result follows.

\blot{
We prove this proposition: Suppose we have a cumulative density function (CDF), $\Phi(x)=0$ (say) if $x<0$. Then we have
 \begin{align*}
\Phi(x)
 & = Prob[x_1 \leq x] \\
 & = Prob[a(1+x_1) \leq x]  \\
 & = E[ \ Prob[a(1+x_1) \leq x \ | \ a] \ ]  \\
 & = \int_0^1 Prob[a(1+x_1) \leq x] f(a) da  \\
 & = \int_0^1 Prob[x_1 \leq x/a -1] f(a) da \\
 & = \int_0^1 \Phi(x/a -1)H(x/a -1) f(a) da, 
\end{align*}
where $H(\cdot)$ is the Heaviside function since $\Phi(x_1)=Prob[x_1 \le a]$ and since $\Phi(x)=0$ if $x<0$. 
Hence the CDF satisfies

\begin{equation}
 \label{eq_IE_CDF}
 \Phi(x) = \int_0^{\min(x,1)} \Phi \left( \frac{x}{a} - 1 \right) f(a)da.
\end{equation}

Now, since the probability density function is the derivative of the CDF, we would like to differentiate $\Phi(x)$ 
with respect to z and use the Leibniz rule. However, it is not clear whether $\Phi(x)$ is differentiable or not. 
Since we know that $f(a)$ is differentiable, we change variables to $w=x/a-1$ to obtain

\begin{displaymath}
 \Phi(x) = x \int_{\max(0,x-1)}^{\infty} \Phi(w) f\left( \frac{x}{w+1} \right) \frac{1}{(w+1)^2} \ dw.
\end{displaymath}
We can differentiate the right-hand side of the above equation for all $z$ and so we can also differentiate 
the left-hand side. Having established that we can differentiate $\Phi(x)$ with respect to $x$, we return 
to (\ref{eq_IE_CDF}). However, it is not clear whether equation~(\ref{eq_IE_CDF}) is differentiable at $x=1$. 
So, we carefully consider two cases $x<1$ and $x>1$. In the case of $x<1$, we obtain

\begin{displaymath}
 \Phi(x) = \int_0^x \Phi\left( \frac{x}{a}-1 \right) f(a)da,
\end{displaymath}
and, by differentiation, we also obtain

\begin{displaymath}
 \Phi'(x) = \phi_l(x) = \Phi(0)f(x) + \int_0^x \phi\left( \frac{x}{a}-1 \right) \frac{f(a)}{a} \ da.
\end{displaymath}

Similarly, for the case of $x>1$, we obtain

\begin{displaymath}
 \Phi(x) = \int_0^1 \Phi\left( \frac{x}{a}-1 \right) f(a)da,
\end{displaymath}
and, by differentiation, we also obtain

\begin{displaymath}
 \Phi'(x) = \phi_r(x) = \int_0^1 \phi\left( \frac{x}{a}-1 \right) \frac{f(a)}{a} \ da.
\end{displaymath}

Taking left-sided and right-sided limits, we have

\begin{displaymath}
 \lim_{x \rightarrow 1^-} \phi_l(x) = \int_0^1 \phi\left( \frac{x}{a}-1 \right) \frac{f(a)}{a} \ da,
\end{displaymath}
and

\begin{displaymath}
 \lim_{x \rightarrow 1^+} \phi_r(x) = \int_0^1 \phi\left( \frac{x}{a}-1 \right) \frac{f(a)}{a} \ da.
\end{displaymath}
Thus, by the sandwich theorem, we obtain

\begin{displaymath}
  \phi(x) = \int_0^{\min(x,1)} \phi\left( \frac{x}{a}-1 \right) \frac{f(a)}{a} \ da.
\end{displaymath}
}

\begin{acknowledgments}
K.P.O is supported by the University of Strathclyde. The simulation data were obtained using the Faculty of Engineering 
High Performance Computer at the University of Strathclyde.
\end{acknowledgments}

\bibliography{aipsamp}% Produces the bibliography via BibTeX.

\begin{thebibliography}{9}
\bibitem{comment} The advantage of the MF approach we adopt
  here is that it leads to a closed form for the capture zones in
  Equation~(\ref{eqDFPE_CZD}), whereas the convolution of the GSD does
  not.
\end{thebibliography}

\end{document}